%% file: main.tex
\documentclass[acmtog,nonacm]{acmart}

\usepackage{float}
\usepackage{subcaption}
\usepackage{booktabs}
\usepackage{tabularx}
\usepackage{makecell}
\usepackage{array}
\usepackage{multirow}
\usepackage{pifont}

\definecolor{turquoise}{cmyk}{0.65,0,0.1,0.1}
\definecolor{purple}{rgb}{0.65,0,0.65}
\definecolor{darkgreen}{rgb}{0.0, 0.5, 0.0}
\definecolor{darkred}{rgb}{0.5, 0.0, 0.0}
\definecolor{darkblue}{rgb}{0.0, 0.0, 0.5}
\definecolor{blue}{rgb}{0.0, 0.0, 1.0}
\definecolor{orange}{rgb}{1.0, 0.5, 0.0}
\definecolor{red}{rgb}{1.0, 0.0, 0.0}
\definecolor{cherry}{RGB}{186,12,47}

\AtBeginDocument{%
  }

\newcommand{\ch}[1]{#1}
\newcommand{\cut}[1]{#1}
\begin{document}

\title{Greenscapes: Optimized Vegetation Configurations in Cities}

\author{Patryk Bartkowiak}
\orcid{0009-0005-6659-8302}
\affiliation{%
  \institution{Adam Mickiewicz University, Poznań}
  \city{Poznań}
  \country{Poland}}
\email{patryk.bartkowiak@amu.edu.pl}

\author{Marcin Marlewski}
\orcid{0009-0008-5434-6388}
\affiliation{%
  \institution{Adam Mickiewicz University, Poznań}
  \city{Poznań}
  \country{Poland}}
\email{marmar73@st.amu.edu.pl}

\author{Paweł Wójcik}
\orcid{0009-0001-9660-0654}
\affiliation{%
  \institution{Adam Mickiewicz University, Poznań}
  \city{Poznań}
  \country{Poland}}
\email{pawwoj10@st.amu.edu.pl}

\author{Jae Joong Lee}
\orcid{0000-0002-0445-3141}
\affiliation{%
  \institution{Purdue University}
  \city{West Lafayette}
  \country{United States of America}}
\email{lee2161@purdue.edu}

\author{Dominik L. Michels}
\orcid{0000-0002-1621-325X}
\affiliation{%
  \institution{King Abdullah University of Science and Technology (KAUST)}
  \city{Thuwal}
  \country{Saudi Arabia}}
\email{dominik.michels@kaust.edu.sa}

\author{Bedrich Benes}
\orcid{0000-0002-5293-2112}
\affiliation{%
  \institution{Purdue University}
  \city{West Lafayette}
  \country{United States of America}}
\email{bbenes@purdue.edu}

\author{Sören Pirk}
\orcid{0000-0003-1937-9797}
\affiliation{%
  \institution{Christian-Albrechts-Universität zu Kiel}
  \city{Kiel}
  \country{Germany}}
\email{soeren.pirk@gmail.com}

\author{Wojciech Pałubicki}
\orcid{0000-0002-2374-346X}
\affiliation{%
  \institution{Adam Mickiewicz University, Poznań}
  \city{Poznań}
  \country{Poland}}
\email{wojciech.palubicki@amu.edu.pl}

\renewcommand{\shortauthors}{Bartkowiak et al.}

\input{src/00-abstract}

\begin{CCSXML}
<ccs2012>
   <concept>
       <concept_id>10010147.10010371.10010396.10010398</concept_id>
       <concept_desc>Computing methodologies~Mesh geometry models</concept_desc>
       <concept_significance>500</concept_significance>
       </concept>
   <concept>
       <concept_id>10010147.10010257.10010293.10010294</concept_id>
       <concept_desc>Computing methodologies~Neural networks</concept_desc>
       <concept_significance>500</concept_significance>
       </concept>
 </ccs2012>
\end{CCSXML}

\ccsdesc[500]{Computing methodologies~Mesh geometry models}
\ccsdesc[500]{Computing methodologies~Neural networks}

\keywords{Urban Modeling, Vision-Language-Models, Scene Generation, Modeling of Parks, Tree Modeling}

\begin{teaserfigure}
  \centering
  \includegraphics[width=1.0\textwidth]{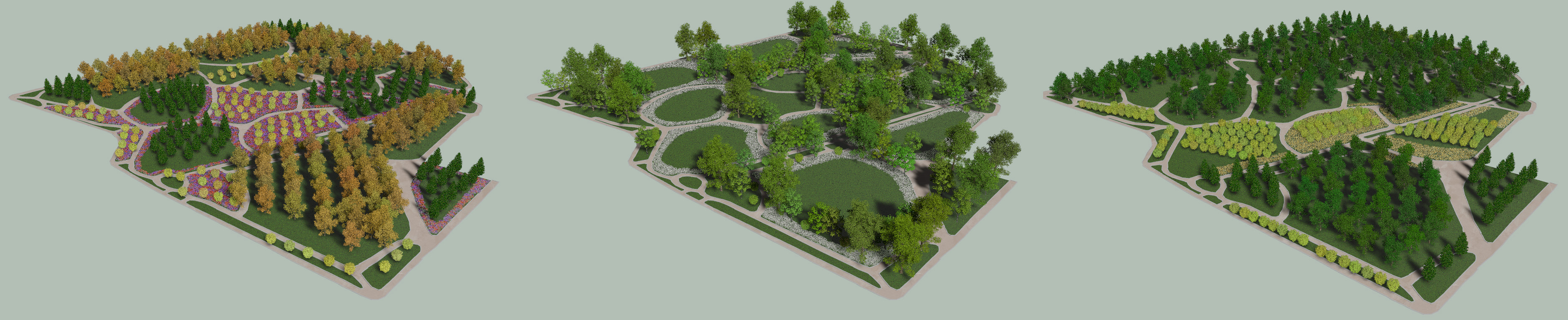}
  \caption{Three complete park designs synthesized from identical Open Street Map site constraints but different style prompts. Left: \emph{formal} design and regular spacing. Middle: \emph{picturesque}-asymmetric composition and clustered planting. Right: \emph{modernist}-clear geometry. All scenes are generated 3D layouts optimized under perceptual and functional objectives using a non-convex approach.}
  \vspace{6mm}
  \label{fig:teaser}
\end{teaserfigure}

\maketitle

\input{src/01-introduction}
\input{src/02-related_work}
\input{src/04-method}
\input{src/05-results}
\input{src/06-conclusion}

\input{src/07-figures}

\begin{acks}
This work is supported by ERC grant \#101170158 - WildfireTwins. This work was supported in part by the U.S. National Science Foundation under awards No. 2412928, 2506783, and 2417510. Any opinions, findings, and conclusions or recommendations expressed in this material are those of the author(s) and do not necessarily reflect those of the National Science Foundation. This work is based upon efforts supported by the USDA-NIFA grants 2024-67021-42879 ``Improving Forest Management Through Automated Measurement of Tree Geometry'' and 2023-08563 ``Optimizing soybean canopy architecture for efficient light and water use under climate change''. The views and conclusions contained herein are those of the authors and should not be interpreted as representing the official policies, either expressed or implied, of the U.S. Government or NRCS. The U.S. Government is authorized to reproduce and distribute reprints for governmental purposes notwithstanding any copyright annotation therein.

\end{acks}

\bibliographystyle{ACM-Reference-Format}
\bibliography{main}

\end{document}

\title[GreenScapes Supplementary Material]{GreenScapes: Optimized Vegetation Configurations in Cities\\Supplementary Material}

\author{Patryk Bartkowiak}
\affiliation{\institution{Adam Mickiewicz University}\city{Pozna\'n}\country{Poland}}
\author{Marcin Marlewski}
\affiliation{\institution{Adam Mickiewicz University}\city{Pozna\'n}\country{Poland}}
\author{Pawe\l{} W\'ojcik}
\affiliation{\institution{Adam Mickiewicz University}\city{Pozna\'n}\country{Poland}}
\author{Jae Joong Lee}
\affiliation{\institution{Purdue University}\country{United States of America}}
\author{Dominik L. Michels}
\affiliation{\institution{King Abdullah University of Science and Technology (KAUST)}\country{Saudi Arabia}}
\author{Bedrich Benes}
\affiliation{\institution{Purdue University}\country{United States of America}}
\author{S\"oren Pirk}
\affiliation{\institution{Christian-Albrechts-Universit\"at zu Kiel}\country{Germany}}
\author{Wojciech Pa\l{}ubicki}
\affiliation{\institution{Adam Mickiewicz University}\city{Pozna\'n}\country{Poland}}
\renewcommand{\shortauthors}{Bartkowiak et al.}

\begin{CCSXML}
<ccs2012>
   <concept>
       <concept_id>10010147.10010371.10010352.10010379</concept_id>
       <concept_desc>Computing methodologies~Physical simulation</concept_desc>
       <concept_significance>300</concept_significance>
       </concept>
   <concept>
       <concept_id>10010147.10010371.10010396</concept_id>
       <concept_desc>Computing methodologies~Computer graphics</concept_desc>
       <concept_significance>300</concept_significance>
       </concept>
</ccs2012>
\end{CCSXML}
\ccsdesc[300]{Computing methodologies~Physical simulation}
\ccsdesc[300]{Computing methodologies~Computer graphics}
\keywords{urban vegetation, landscape modeling, vision-language models, CMA-ES, procedural modeling}

\maketitle

\section{Overview}
This supplementary material provides implementation and reproducibility details for the GreenScapes pipeline. The main paper describes the complete method: an OSM-derived site representation is converted into a top-down conditioning image; a VLM produces a semantic vegetation-layout prior; detected instances are projected into a metric 3D scene and repaired if needed; and CMA-ES refines the resulting configuration under perceptual and functional objectives. This document provides the VLM protocol, deterministic post-processing and repair rules, CMA-ES settings, objective configurations, optimization traces, and acoustic-exposure parameters used in the reported experiments. \ch{All settings reported here correspond to the released implementation (Sec.~\ref{sec:supp_release}).}

\section{VLM Prompting and Model Configuration}
\label{sec:supp_vlm}

\subsection{Model and API settings}
All reported VLM layout-generation experiments use the Google Gemini API model string \texttt{gemini-3-pro-image-preview}. We refer to this model as the Gemini~3 Pro image preview in the paper. The image-editing stage uses a fixed prompt template and a site-specific conditioning image. Unless stated otherwise, the generated image resolution is \(1024\times1024\) with aspect ratio \(1{:}1\). Each prompt produces one accepted image; failed API calls are retried up to five times using the same prompt and input image. The object-annotation pass uses the same model string and a fixed JSON/bounding-box instruction. \ch{For the JSON box-annotation call, the temperature is set to \(0.5\); the dot-annotation variant used for the city-scale examples uses the API's default sampling parameters.} The image-generation call uses the API's image-generation defaults for sampling parameters that are not explicitly exposed in our scripts. \ch{Note that the Gemini image-generation API does not expose a random seed; VLM samples are therefore stochastic, and all reported robustness statistics are computed over independently sampled layouts. All downstream pipeline stages use fixed seeds.}

\ch{In addition to the layout-generation and annotation prompts, the pipeline uses three fixed viewpoint-annotation prompts: one that marks pedestrian standing points, one that marks points of visual interest, and one that connects standing points to interest points. The annotated rasters are decoded deterministically by color thresholding and connected-component analysis into view positions and directions. These prompts are included verbatim in the released code.}

\begin{table}[t]
\centering
\caption{VLM settings used for the reported experiments. ``Prompt instance'' denotes a fixed filled prompt, i.e., a template together with the corresponding arrangement rules.}
\label{tab:vlm_settings}
\small
\begin{tabularx}{\linewidth}{@{}lY@{}}
\toprule
Setting & Value \\
\midrule
VLM model/version & \texttt{gemini-3-pro-image-preview} \\
Layout-generation input & OSM-derived top-down conditioning image plus fixed filled prompt \\
Layout-generation output & Complete edited top-down image with vegetation added only on grass/editable pixels \\
Image resolution & \(1024\times1024\) \\
Aspect ratio & \(1{:}1\) \\
Accepted candidates & One accepted generated image per prompt instance \\
Retry budget & Five API retries for failed image-generation calls \\
Annotation model & \texttt{gemini-3-pro-image-preview} \\
Annotation output & JSON bounding boxes with labels \texttt{bigtree}, \texttt{smalltree}, \texttt{shrub} \\
Annotation temperature & \ch{\(0.5\) (JSON box annotation); API defaults for the dot-annotation variant} \\
Prompt protocol & Fixed templates; no interactive prompt tuning during the reported experiments \\
Reported prompt set & 18 fixed prompt instances: 17 layout/style/placement instances and one object-annotation instance\ch{; three auxiliary viewpoint-annotation prompts (all released with the code)} \\
\bottomrule
\end{tabularx}
\end{table}

\subsection{Representative layout-generation prompt}
\label{sec:supp_prompt_template}
The following listing is a representative park-generation template used by the VLM stage. The placeholder \texttt{\{user\_query\}} is filled with a fixed arrangement-rule block from a YAML configuration file. The purpose of the template is to constrain the VLM to a strict top-down image-editing task: non-grass pixels are locked, the input geometry is preserved, and vegetation is placed only on grass pixels. \ch{The complete set of templates and arrangement-rule blocks is included in the released code.}

\begin{lstlisting}[style=promptstyle,caption={Representative park-generation prompt template used for top-down VLM layout generation.},label={lst:park_prompt}]
### ROLE
You are a high-end landscape architecture engine. Your task is to perform a STRICT IMAGE EDIT on the input: keep the park exactly the same and ONLY add plants on the grass. YOU MUST OUTPUT A COMPLETE IMAGE as your response - this is an image generation task, not a text description task.

### INPUT CONTEXT
- The input image is a top-down orthographic view of a large park area.
- Grey strip = The Path (OBSTACLE - DO NOT CHANGE).
- Blue areas (ponds/lakes) = Water (OBSTACLE - DO NOT CHANGE).
- Dark Green area = The Grass/Lawn (TARGET AREA FOR VEGETATION).
- This is a park setting with open green spaces, paths, and water.
- The light gray area outside the park silhouette is BACKGROUND/EMPTY CANVAS (OUT OF SCOPE - DO NOT CHANGE).

### VISUAL STYLE & FORMAT
- *Perspective:* Strictly maintain the 2D top-down orthographic view.
- *Render Style:* Photorealistic aerial view. Plants must have realistic texture (leaves/branches).
- *Shadows:* Soft, natural top-down shadows are permitted on the grass only (adds realism/depth).

### ASSET LIBRARY
You are ONLY allowed to use these 3 specific asset classes. ENSURE HIGH COLOR CONTRAST between plants and the grass so they are easily detected:
1. *Big Tree:* Large, DEEP FOREST GREEN, textured canopy (circular shape). Must appear significantly DARKER than the grass.
2. *Small Tree:* Medium, BRIGHT LIME-GREEN or GOLDEN-GREEN, distinct canopy (circular shape). Must appear significantly LIGHTER/BRIGHTER than the grass.
3. *Shrub:* Small, compact, BLUEISH bush (individual circular plant, not a hedge). Must appear BLUE/BLUE-GREEN to be visually DISTINCT from the lawn.

### CRITICAL CONSTRAINTS (STRICT ENFORCEMENT)
1. *STRICT IMAGE EDIT (HIGHEST PRIORITY):* Treat the input as the final ground truth. You MUST keep the image geometry and layout EXACTLY the same.
2. *BOUNDARIES & GEOMETRY LOCK:* The park outline, edges, corners, boundaries, and the path geometry/width/position are FIXED. Do NOT move, warp, rotate, resize, crop, or redraw anything.
3. *CANVAS / SILHOUETTE LOCK (VERY IMPORTANT):* The park has a specific irregular silhouette. Do NOT extend it, square it off, or fill the surrounding canvas. The outside background must remain exactly as in the input (same color/texture).
4. *PIXEL IDENTITY RULE:* Every pixel that is NOT grass (paths, water, edges, boundaries, outside background/canvas, any non-grass features) must remain IDENTICAL to the input image. Do not repaint, recolor, relight, blur, sharpen, or change textures outside the grass.
5. *EDIT SCOPE LIMIT:* The ONLY allowed changes are: (a) adding plant canopies on grass pixels and (b) optional soft top-down shadows cast onto grass pixels. Nothing else.
6. *NO BACKGROUND PAINTING:* Do NOT add grass, trees, shrubs, shadows, haze, vignette, or any marks to the outside background/canvas. Do NOT create a rectangular frame.
7. *INFRASTRUCTURE FREEZE:* The existing path must NOT be extended, changed, widened, narrowed, recolored, or added to. No new paths, fences, curbs, or paved areas.
8. *MATERIAL PURITY:* The grass zone is for VEGETATION ONLY. It is strictly forbidden to place gravel, stone, concrete, mulch, dirt, or any new ground material on the grass.
9. *ASSET ISOLATION (MANDATORY):* Every single plant MUST be an individual, distinct CIRCULAR shape surrounded by a visible ring of grass.
10. *NO HEDGES OR CLUMPS:* Do NOT create continuous hedges, merged clumps, or elongated shapes. Even when creating "lines" or "walls" of plants, they must be composed of separate, individual circular dots (e.g., o o o o o), not a solid line (-----).
11. *MINIMUM GAP FOR DETECTION (VERY IMPORTANT):* Keep a clearly visible grass gap between any two plants. The gap must be large enough to be obvious at a glance (prefer fewer plants rather than smaller gaps). Aim for at least ~5 pixels of grass between canopies at 1024x1024 resolution, scaled proportionally for other resolutions.
12. *CLUSTER SEPARATION:* Even when creating "dense clusters", "groves", or "forests", you MUST keep a visible gap between every single tree. A "grove" is a collection of separate trees, not a single blob. If you cannot fit them with gaps, place fewer trees.
13. *MAXIMUM PLANT COUNT:* Do not overcrowd the park. Limit the total number of plants to approximately 300. Prioritize clear separation over quantity.
14. *MINIMUM PLANT SIZE:* Do not generate tiny, microscopic specks. Every plant must be large enough to be clearly identified as a distinct object.
15. *FAILSAFE:* If a requested change would require modifying geometry/boundaries/path/outside background or any non-grass pixels, DO NOT do it. Leave that area unchanged and only place plants where it is safe.

### ARRANGEMENT INSTRUCTIONS
Apply the following rules, strictly adhering to the "Asset Isolation" constraint:

"{user_query}"

### OUTPUT COMMAND
YOU MUST OUTPUT A COMPLETE IMAGE. Generate and return the modified image as your response. The output MUST be an image file showing the park with plants added to the grass areas. Preserve the park silhouette/geometry/boundaries exactly as in the input (including the outside background), and ONLY populate the existing grass with plants (with visible grass gaps between all canopies). Do not return text-only responses - you must generate and output the image.
\end{lstlisting}

\subsection{Representative arrangement-rule block}
The next listing gives one example of the arrangement rules inserted into the template above. Other reported style and placement prompts use the same template structure but substitute different fixed arrangement-rule blocks.

\begin{lstlisting}[style=promptstyle,caption={Representative fixed arrangement-rule block for a French formal garden prompt.},label={lst:french_rules}]
- STYLE: French Formal Garden (Jardin a la francaise) - Focus on symmetry, geometry, and strict proportion.
- GEOMETRY: Ignore natural scattering. Arrange all vegetation in strict geometric patterns (lines, grids, rectangles, circles).
- SEPARATION: STRICTLY ENFORCE gaps between every plant. Do not create solid lines. Every geometric shape must be made of individual, distinct dots (plants).
- AXIS & ALIGNMENT: Use the path as a central axis. Align trees in parallel rows (allees) along the path edges to create a grand perspective.
- BOSQUETS (GROVES): On larger grass areas, arrange Big Trees in perfect square or rectangular grids with clear spacing between canopies.
- PARTERRES: Use Shrubs to create geometric outlines (rectangles, circles) on open grass. IMPORTANT: These outlines must be dotted lines of separate shrubs, NOT continuous hedges.
- BOUNDARIES: Use Small Trees or rows of separate Shrubs to define outdoor rooms. Keep them distinct.
- SYMMETRY: Attempt to mirror the planting arrangement across the central axis where the path geometry allows.
- NO RANDOMNESS: Do not place any single isolated plant randomly. Every plant must be part of a larger geometric unit or line.
\end{lstlisting}

\subsection{Object-annotation prompt}
The VLM annotation pass converts the generated top-down image into a list of object boxes. Each returned object is required to use one of the three vegetation labels used by the downstream pipeline. Boxes are returned in normalized image coordinates and then converted into pixel coordinates and metric scene coordinates. \ch{For the city-scale examples, a dot-annotation variant of this protocol is used instead: the model stamps color-coded dots (one per plant), which are extracted by per-channel color thresholding and connected-component analysis.}

\begin{lstlisting}[style=promptstyle,caption={Representative object-annotation prompt for extracting individual vegetation instances.},label={lst:annotation_prompt}]
Detect all plants (trees, shrubs, bushes) in this garden image. Return bounding boxes for each INDIVIDUAL plant. IMPORTANT: Plants are often circular. If you see a large irregular or elongated shape, it is likely a cluster of multiple plants - break it down into separate circular bounding boxes. Do not group multiple plants into one large box. CRITICAL: Only detect plants INSIDE the park/parcel boundaries (the grass areas). Ignore any vegetation or background elements outside the park shape. You MUST label each plant with exactly one of these three labels based on its appearance and COLOR:
- 'bigtree': DEEP FOREST GREEN / DARK GREEN.
- 'smalltree': BRIGHT LIME-GREEN / LIGHT GREEN.
- 'shrub': BLUE / BLUE-GREEN bush.
Do not use any other labels or descriptive text.

System instruction: Return bounding boxes as a JSON array with labels. Never return masks or code fencing. Each label MUST be exactly one of: 'bigtree', 'smalltree', or 'shrub'. Use 'bigtree' for large trees, 'smalltree' for medium-sized trees, and 'shrub' for small bushes/shrubs. Classify based on the visual definitions: Big Tree = Dark Green, Small Tree = Light/Lime Green, Shrub = Blue/Blue-Green. CRITICAL: Avoid merging clusters. Assume plants are distinct circular canopies. If shapes overlap, detect them as separate objects. Strictly IGNORE anything outside the main park/parcel area. Limit detection to the most prominent ~300 plants if there are too many.
\end{lstlisting}

\section{Projection, Feasibility Checks, and Repair}
\label{sec:supp_repair}

Each VLM bounding box is assigned to a vegetation class and mapped from image coordinates to lot coordinates using the inverse \ch{affine transform} \(T^{-1}\) defined by the top-down rasterization. The box center defines the initial 2D instance position, and the box extent defines the initial footprint used to instantiate a 3D bounding ellipsoid. \ch{Feasibility tests operate on the 2D footprint disc of each instance (the equatorial cross-section of its bounding ellipsoid); vertical extents are assigned from class- and asset-specific priors.} The instance is invalid if its center lies outside the editable planting mask, if its \ch{footprint} extends outside the lot, if it intersects obstacle polygons, if it overlaps path geometry, or if it violates vegetation--vegetation collision constraints.

Invalid instances are repaired by a deterministic two-stage operator. First, center reprojection moves the center to the nearest feasible location on the planting mask. Nearest feasible location is measured by Euclidean distance in the metric lot coordinate frame. Second, obstacle correction shifts the \ch{footprint} locally along the signed-distance gradient to remove remaining intersections. The maximum number of repair iterations is 30. If the instance is still invalid after 30 iterations, it is discarded. Discarded instances are not counted in reported vegetation-instance counts, biomass calculations, optimization variables, or validation statistics. \ch{The repair operator runs once, before optimization; during CMA-ES refinement, feasibility is maintained by clamping the scale parameters to their bounds (Sec.~\ref{sec:supp_cmaes}).}

\begin{algorithm}[t]
\caption{Deterministic instance projection and repair.}
\label{alg:repair}
\DontPrintSemicolon
\KwIn{VLM boxes \(B\), labels \(c\), image-to-world transform \(T^{-1}\), planting mask \(M\), obstacle/path polygons \(O\), maximum repair iterations \(K=30\).}
\KwOut{Feasible vegetation-instance set \(\Pi\).}
\(\Pi \leftarrow \emptyset\)\;
\ForEach{box \(b_i\in B\) with label \(c_i\)}{
  Convert the box center to world coordinates: \(p_i \leftarrow T^{-1}(\mathrm{center}(b_i))\)\;
  Convert the box extent to a metric footprint and instantiate ellipsoid \(E_i\)\;
  Assign the initial asset category from \(c_i\)\;
  \If{\(p_i\notin M\)}{
    Reproject \(p_i\) to the nearest feasible point in \(M\) using Euclidean distance\;
  }
  \For{\(k=1\) \KwTo \(K\)}{
    \If{\(E_i\) is inside the lot and intersects neither \(O\) nor accepted instances in \(\Pi\)}{
      add instance \((p_i,E_i,c_i)\) to \(\Pi\) and stop repair for this instance\;
    }
    Shift \(p_i\) and \(E_i\) by the local signed-distance correction away from the active violation\;
  }
  \If{no feasible placement was found after \(K\) iterations}{
    discard the instance and do not include it in downstream counts or optimization\;
  }
}
\Return \(\Pi\)\;
\end{algorithm}

\section{CMA-ES Reproducibility Settings}
\label{sec:supp_cmaes}
The optimization stage uses the same \ch{parameterization scheme} across all experiments. Let \(n\) denote the dimensionality of the active optimization vector\ch{: for a scene with $M$ feasible instances, $n=2M$ scale variables, plus $6M$ whitened attribute variables when GMM-based species substitution is enabled. All feasible instances of the repaired configuration are included in the optimization; positions are not optimization variables.} The CMA-ES population size is
\begin{equation}
\lambda=\max\left(4,\left\lfloor 4+3\ln n\right\rfloor\right),
\end{equation}
\ch{and the initial step size $\sigma_0$ is configured per experiment (between $0.15$ and $0.4$ in normalized parameter space for the reported runs). Continuous search variables are unbounded and mapped through a sigmoid to per-instance scale multipliers within a configured range (typically $[0.3,2.0]$ times the default asset size for the reported experiments), followed by absolute clamps on the resulting crown radii and heights. We use a separable (diagonal-covariance) CMA-ES variant; the distribution mean is re-anchored to the best-so-far solution after each generation, and each run starts with a short uniform-exploration stage before per-axis adaptation.} Feasibility is enforced by the repair and rejection rules described in \Cref{sec:supp_repair}.

\ch{The GMM asset prior consists of 22 species kernels with diagonal covariances and uniform mixture weights over the 6D attribute space of the main paper; the full parameter set is released in \texttt{data/species/species.json}. The Mahalanobis admissibility threshold for species substitution is $3.0$; proposals inadmissible for all species are projected back onto the closest kernel's threshold ellipsoid.}

\begin{table*}[t]
\centering
\caption{Reproducibility table for the reported GreenScapes experiments. Values are fixed unless explicitly stated by a specific experiment. \ch{Per-experiment weights, targets, and step sizes are given in the released configuration files (\texttt{configs/optimizer/}).}}
\label{tab:reproducibility}
\small
\begin{tabularx}{\textwidth}{@{}lY@{}}
\toprule
Parameter & Setting \\
\midrule
VLM model/version & \texttt{gemini-3-pro-image-preview} \\
VLM output resolution & \(1024\times1024\), aspect ratio \(1{:}1\) \\
Prompt protocol & Fixed prompt instances; no interactive prompt tuning during reported runs \\
Accepted VLM candidates & One accepted image per prompt instance; failed API calls retried up to five times\ch{; the image API exposes no random seed} \\
Annotation labels & \texttt{bigtree}, \texttt{smalltree}, \texttt{shrub} \\
Maximum generated plant count requested in prompt & Approximately 300 plants for park prompts \\
Image-to-world mapping & Single global \ch{axis-aligned affine} transform \(T\) from metric lot coordinates to image coordinates; detections use \(T^{-1}\) \\
Nearest feasible location metric & Euclidean distance in metric lot coordinates \\
Maximum repair iterations & 30 \\
Discarded invalid instances & Not counted in reported instance counts, biomass, optimization variables, or validation statistics \\
\ch{Evaluation viewpoints} & \ch{VLM-annotated standing points and view directions, decoded by color thresholding; fixed during optimization; view files released with the code} \\
CMA-ES seed & 0 \\
CMA-ES population size & \(\lambda=\max(4,\lfloor 4+3\ln n\rfloor)\) \\
CMA-ES initial step size & \ch{\(\sigma_0\in[0.15,0.4]\) in normalized parameter space, configured per experiment} \\
\ch{CMA-ES variant} & \ch{Separable (diagonal-covariance); mean re-anchored to best-so-far each generation; short uniform-exploration stage at the start of each run} \\
Continuous optimization bounds & \ch{Unbounded search variables mapped through a sigmoid to configured multiplier ranges, followed by absolute clamps on radii and heights} \\
Physical scale range & \ch{Typically \([0.3,2.0]\) times the default asset size (multiplier range), with absolute clamps per experiment} \\
Asset substitution & Deterministic maximum-posterior assignment under the GMM asset prior, followed by admissibility checking\ch{; Mahalanobis threshold $3.0$; 22 species kernels with diagonal covariances (released)} \\
Voxel resolution for acoustic exposure & \(1\,\mathrm{m}\) \\
Acoustic rays per asset & \ch{20{,}000 single-bounce Monte Carlo rays for each of 16 incident directions (RNG seed 12)} \\
Acoustic incident-direction sampling & Cosine-weighted sampling over a hemisphere \\
Acoustic receiver viewpoints & Eight receiver \ch{points at $2\,\mathrm{m}$ ear height, evenly spaced} along the path network \\
Optimization generations & 100 generations for a typical full run; the Joyce Kilmer trace in the main paper visualizes 30 generations \\
Runtime hardware & Windows 11 laptop, Intel Core i9-14900HX CPU, 32 GB RAM, NVIDIA GeForce RTX 4080 mobile GPU \\
Typical runtime & Approximately 23 seconds per generation and approximately 38 minutes for a 100-generation run, excluding VLM API latency \\
\bottomrule
\end{tabularx}
\end{table*}

\section{Objective Terms and Ablation Configurations}
\label{sec:supp_objectives}

Each candidate layout \(R(\theta)\) is evaluated by a weighted objective
\begin{equation}
J(\theta)=
\sum_{k\in\mathcal{K}}w_k\,\widehat{J}_k(R(\theta))
+ w_{\mathrm{reg}}J_{\mathrm{reg}}(\theta)
+ J_{\mathrm{feas}}(\theta),
\end{equation}
where \(\widehat{J}_k\) are normalized objective terms, \(J_{\mathrm{reg}}\) penalizes unnecessary deviations from the initial scale configuration when used, and \(J_{\mathrm{feas}}\) rejects or repairs infeasible candidates. The objective set \(\mathcal{K}\) contains sky-view factor, skyline fit, path shading, acoustic exposure, and biomass terms. \ch{Concretely, each term contributes $w_k\big((v_k-t_k)/\max(|t_k|,\epsilon)\big)^2$, i.e., the squared relative error of the measured value $v_k$ with respect to the target $t_k$, using the target magnitude as the normalization denominator (with fallback denominators of $1.0$, or the rendered image height for the skyline-fit term, when the target is zero). View-dependent terms are averaged over the evaluation viewpoints before weighting. The regularizer $J_{\mathrm{reg}}$ penalizes the mean squared deviation of the scale multipliers from one, with weight $10^{-3}$ unless stated otherwise.} For monotonic terms, the value is normalized over the valid range used by the experiment and assigned the sign required by the optimization direction.

\ch{The skyline-fit term is instantiated per experiment as one of: skyline flatness, skyline periodicity, or the deviation of the extracted skyline profile from a target profile (e.g., a sinusoid). The skyline descriptor comprises two components (flatness and periodicity); counting these components separately yields six metric values over the five objective terms.}

\begin{table*}[t]
\centering
\caption{Objective configurations used for the objective-ablation figure in the main paper \ch{(Fig.~12)}. Each ablation assigns one dominant weight \(\domw\) to the selected term and \ch{low or zero weights} \(\loww\) to the remaining terms\ch{. The exact weight vectors and targets for each panel are provided in the released configuration files (\texttt{configs/optimizer/})}. This preserves weak regularization from the other objectives while visualizing the qualitative effect of the dominant criterion.}
\label{tab:objective_configs}
\small
\begin{tabularx}{\textwidth}{@{}l l Y Y Y@{}}
\toprule
Main-paper panel & Configuration & Dominant term & Optimization direction & Remaining objective weights \\
\midrule
\ch{Fig.~12(b)} & \ch{Skyline sine-function criterion} & \ch{\(\widehat{J}_{\mathrm{sky}}\)} & \ch{Minimize deviation from the target sinusoidal skyline profile} & \ch{SVF, path shading, acoustic exposure, and biomass use \(\loww\)} \\
\ch{Fig.~12(c)} & \ch{SVF ratio} & \ch{\(\widehat{J}_{\mathrm{SVF}}\)} & \ch{Match the target sky-view factor / increase openness when the target requires higher SVF} & \ch{Skyline, path shading, acoustic exposure, and biomass use \(\loww\)} \\
\ch{Fig.~12(d)} & Minimal path shading & \(\widehat{J}_{\mathrm{shade}}\) & Minimize path-level shading according to the ablation target & SVF, skyline, acoustic exposure, and biomass use \(\loww\) \\
\ch{Fig.~12(e)} & Acoustic objective & \(\widehat{J}_{\mathrm{acoustic}}\) & Reduce acoustic exposure / increase excess attenuation at receiver viewpoints & SVF, skyline, path shading, and biomass use \(\loww\) \\
\ch{Fig.~12(f)} & Maximum biomass & \(\widehat{J}_{\mathrm{bio}}\) & Maximize normalized vegetation biomass under feasibility constraints & SVF, skyline, path shading, and acoustic exposure use \(\loww\) \\
\bottomrule
\end{tabularx}
\end{table*}

\paragraph{SVF objective.}
Sky-view factor is computed from a fisheye semantic segmentation buffer as the fraction of sky pixels inside the circular fisheye domain. \ch{The fisheye view is composited from five cube-map faces using an equidistant hemispherical projection.} The SVF objective is a target-matching term when a target SVF is specified.

\paragraph{Skyline objective.}
The skyline term is computed from \ch{the eye-level perspective segmentation view}. We extract a skyline profile column-wise, detrend it, and compute flatness and periodicity descriptors. The sine-function ablation uses a target skyline profile and penalizes deviations from it.

\paragraph{Path-shading objective.}
The path-shading term is computed over samples of walkable paths using the deterministic shadow-propagation approximation described in the main paper. Depending on the experiment, this term can be used either as a target-matching term or as a monotonic term. The ablation in \cref{tab:objective_configs} uses the minimal path-shading configuration.

\paragraph{Acoustic-exposure objective.}
The acoustic term is computed as the excess attenuation relative to a free-field reference, with fixed source, receiver, grid, timestep, and boundary settings. It is treated as a deterministic optimization surrogate rather than a calibrated wave-acoustic simulation.

\paragraph{Biomass objective.}
The biomass term is computed from the asset-specific biomass coefficients and current vegetation scale\ch{, i.e., as occupied canopy volume on the voxel grid scaled by the per-species coefficient $k_{\mathrm{bio}}$}. The ablation rewards higher normalized biomass while preserving feasibility constraints.

\section{Optimization Trace}
\label{sec:supp_optimization_trace}

\begin{figure*}[t]
  \centering
  \includegraphics[width=0.82\textwidth]{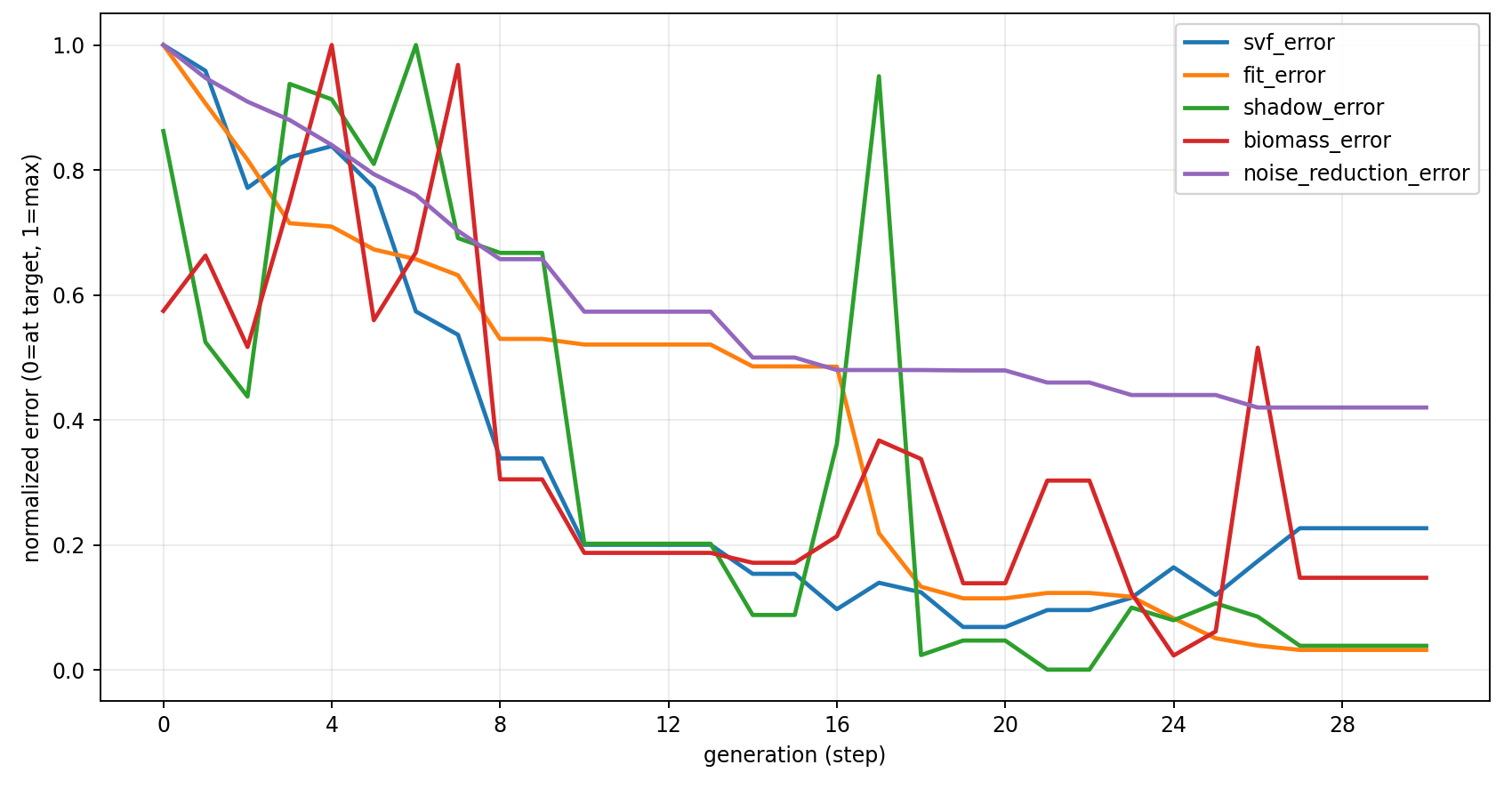}
  \caption{CMA-ES optimization trace for Joyce Kilmer Park. Normalized target error (0 = at target, 1 = worst) over generations for SVF, skyline fit, shade, biomass, and noise reduction. Most terms drop rapidly in the first \(\sim 10\) generations, then refine more slowly with occasional non-monotonic trade-offs. Noise reduction decreases more gradually, indicating a harder objective under the same geometric constraints.}
  \label{fig:supp_combined}
\end{figure*}

Figure~\ref{fig:supp_combined} plots the CMA-ES optimization trace for Joyce Kilmer Park as \emph{normalized target error} (0 = at target, 1 = worst). Starting from the VLM-initialized scene at generation 0, CMA-ES rapidly decreases most errors within the first \(\sim 10\) generations. After this initial drop, the curves enter a slower refinement regime where improvements become incremental and occasionally non-monotonic, most visibly for shade and biomass, reflecting trade-offs introduced by discrete asset substitutions and constraint repairs. By the end of the run (\(\sim 30\) generations), SVF, skyline fit, shade, and biomass converge to low residual error, while noise reduction remains higher, indicating that the acoustic objective is harder to satisfy simultaneously under the same geometric constraints.

\section{Acoustic-Exposure Details}
\label{sec:supp_acoustics}

Acoustic exposure is evaluated with a two-stage surrogate model designed for repeated use inside CMA-ES. In preprocessing, each vegetation asset is assigned three geometric-acoustic descriptors computed from its 3D geometry. \ch{We trace 20{,}000 single-bounce Monte Carlo rays for each of 16 incident directions per asset (RNG seed 12).} Incident directions are sampled over a hemisphere using cosine-weighted sampling. At the first intersection, the outgoing direction is sampled as either specular or diffuse reflection, with diffuse directions drawn from a cosine-weighted lobe. From the resulting outgoing-direction distribution we compute a random-scattering fraction \(s_{\mathrm{rand}}\), a normalized angular-spread descriptor \(d_n\in[0,1]\), and a mean outgoing direction \(\boldsymbol{\mu}\).

For each candidate scene, the park volume is discretized at \(1\,\mathrm{m}\) voxel resolution, and voxels are labeled as air, vegetation, building, or ground. Sound-energy density \(E(\mathbf{x},t)\) is evolved with an anisotropic diffusion-with-decay model,
\begin{equation}
\frac{\partial E}{\partial t}
=\nabla\cdot(D(\mathbf{x})\nabla E)-\lambda(\mathbf{x})E+S(\mathbf{x},t),
\end{equation}
where \(S\) is a point-source term, \(D\) is a material-dependent diffusion tensor, and \(\lambda\) is a material-dependent decay rate. Vegetation voxels receive parameters from the precomputed descriptors using the same global heuristic in all experiments: larger \(d_n\) increases diffusion magnitude, \(\boldsymbol{\mu}\) biases the anisotropy direction, and larger \(s_{\mathrm{rand}}\) increases decay. Buildings are treated as high-decay regions, and air uses fixed background parameters. We use a CFL-stable explicit finite-difference timestep and open top and side boundaries.

\ch{For completeness, the released implementation uses the following fixed mapping constants. Air: diffusion $D_{\mathrm{air}}=0.02$, decay $\lambda_{\mathrm{air}}=0.1$. Vegetation: base diffusion $\min(d_n\cdot 3.0,\;3.0\,D_{\mathrm{air}})$, distributed anisotropically across the coordinate axes by weighting with the normalized components of $\boldsymbol{\mu}$ (anisotropy strength $0.2$, with a lower bound of $0.001$ per axis); decay $\mathrm{clip}(\alpha,0,0.99)\cdot 0.37$, where $\alpha=\mathrm{clip}\big(0.05+0.3\,\overline{s_{\mathrm{rand}}}+0.1\,\mathrm{std}(s_{\mathrm{rand}}),\,0,\,0.99\big)$ aggregates the per-direction scattering fractions. Buildings: $D=0.001$, $\lambda=5.0$. The point source has strength $9\times 10^{15}$; the timestep uses a CFL safety factor of $0.4$; simulations run for at most $60{,}000$ steps or until the total energy falls below a relative threshold of $10^{-14}$. The reported attenuation is the mean over receivers of the time-integrated energy ratio. Note that with the released species parameters, the vegetation base diffusion saturates at the air-derived cap for all species, so species differ acoustically through their anisotropy direction and decay rather than diffusion magnitude.}

Exposure is evaluated at eight receiver viewpoints sampled along the path network. For each candidate scene, we compare the vegetation simulation against a free-field reference with identical source, receiver, grid, timestep, and boundary settings. The reported objective uses excess attenuation,
\begin{equation}
\Delta L_{\mathrm{dB}}=-10\log_{10}\left(\frac{E_{\mathrm{with}}}{E_{\mathrm{free}}}\right),
\end{equation}
averaged over receivers. This gives a deterministic acoustic-exposure term that is comparable across optimization iterations.

To check that the surrogate produces plausible magnitudes, we ran a single-belt sanity test: a linear tree belt of \(12\,\mathrm{m}\) width separates an omnidirectional source from three receivers at \(2\,\mathrm{m}\) ear height. Excess attenuation across receivers falls in the \(5\)--\(10\,\mathrm{dB}\) range, consistent with the \(3\)--\(8\,\mathrm{dB}\) reported by~\citet{AttenboroughTaherzadeh2016Forests} for tree belts of similar width and the ground-attenuation guidance of ISO~9613-2~\ch{\cite{ISO9613Part2}}. The agreement is qualitative, not calibrated, and is reported only to verify that the diffusion-with-decay surrogate stays within the right order of magnitude for the regimes used during optimization. \ch{This validation is included as an automated test in the released code.}

\section{Notes on Reported Counts and Statistics}
\label{sec:supp_counts}

All reported instance counts are computed after VLM annotation, metric reprojection, feasibility checking, and repair. Instances discarded after failed repair are excluded. This convention is used consistently for the large-scale examples, biomass calculations, optimizer variable construction, and validation statistics. The convention avoids counting objects that were proposed by the VLM but could not be placed in the valid metric scene.

Across 147 park-generation experiments, the VLM-derived initial layout contained at least one invalid placement in 48 cases. Such invalid placements were corrected by the deterministic repair stage when possible; otherwise, the offending instances were discarded and excluded from downstream counts. In all 147 experiments, the optimization stage modified the VLM-derived scene through feasible configuration changes, scale adjustment, or asset/species substitution. This supports the interpretation of the VLM as a semantic prior rather than as the final layout generator.

\ch{\section{Code and Data Release}
\label{sec:supp_release}
We release the implementation of the GreenScapes pipeline at \url{https://github.com/greenscapes-project/greenscapes}. The release comprises the optimization engine (Rust), the VLM pipeline and acoustic surrogate (Python), all VLM prompt templates and arrangement-rule blocks, the object- and viewpoint-annotation prompts, the projection and repair operators, the objective definitions and per-experiment optimizer configurations, the species database with the complete GMM parameters (\texttt{data/species/species.json}), the acoustic descriptor precomputation and diffusion solver, analysis tooling (style metrics, optimization traces, attribute-space visualization), site data for the reported examples, and automated tests, including the acoustic tree-belt validation. Because the commercial 3D vegetation assets used for the paper's renderings cannot be redistributed, the release includes a script that generates placeholder assets with identical bounding volumes, so quantitative results reproduce while renderings appear schematic. Because VLM APIs may change over time, the release records the exact model string (\texttt{gemini-3-pro-image-preview}), prompt instances, image resolution, and post-processing rules used for the reported experiments; the image-generation API exposes no random seed, so VLM outputs are stochastic.}

\bibliographystyle{ACM-Reference-Format}
\bibliography{main}

%% file: src/00-abstract.tex
\begin{abstract}
Vegetation plays a central role in cities, serving functional roles such as balancing shade and sky exposure, shaping the microclimate, carbon sequestration, and providing acoustic screening. Furthermore, aesthetic choices in landscape styles (formal, modernist, picturesque) are important for citizens' well-being and should be considered in the design of urban landscapes. In this paper, we present a novel method for placing vegetation \ch{-- ranging from shrubs to large trees --} in 3D urban scenes. Our method uses individual lots as input and employs a conditioned Vision-Language Model (VLM) to generate a top-down rendering of vegetation\ch{, placing large trees, small trees, and shrubs on lawn areas}. This rendering is then converted into a 3D scene by detecting individual plants and selecting appropriate 3D assets. \ch{To refine the proposed configurations, we define an optimization loop that keeps the VLM-generated positions fixed after a deterministic feasibility repair and uses CMA-ES to optimize per-instance scale transformations along with discrete vegetation asset substitutions.} To define the objective function for the optimization, \ch{we evaluate the generated 3D scenes with efficient surrogate models of shade and path-based acoustic exposure}. Furthermore, we render in-scene views at each optimization step to compute view-dependent perceptual descriptors, enabling the generation of both aesthetic and functional vegetation layouts for urban scenes.
\end{abstract}

%% file: src/01-introduction.tex
\section{Introduction}
Synthesizing urban vegetation in computer graphics poses substantial challenges due to the interplay of planning constraints, aesthetic composition across multiple scales, from individual plants to city-wide ecosystems, and vegetation diversity. Modeling urban landscapes is critical for visual storytelling in games and movies~\cite{smelik2014survey}, urban planning~\cite{Vanegas:2012}, microclimate analysis~\cite{toparlar2017review}, and digital twins~\cite{schrotter2020digital}. Moreover, urban afforestation plays a critical role in making cities more livable~\cite{EU_NatureRestorationReg_2024_1991, EC_3BillionTrees_Factsheet}. 

While computer graphics has made significant progress in procedural city generation and building synthesis~\cite{parish2001procedural,mueller2006procedural,nishida2016interactive}, most existing techniques either reduce vegetation to stochastic point distributions or focus solely on trees, ignoring the understory, shrubbery, and ground cover. Ecosystem simulations provide high-fidelity growth models but are designed for wild, competitive environments~\cite{10.1145/3528223.3530146,10.1145/3641816,Gain-Ecobrush,10.1145/2766975} and not for managed urban spaces~\cite{10.1145/1944745.1944773}. Others focus on learning distributions from satellite data~\cite{niese2022procedural}, but often lack the precise controllability required to enforce specific setbacks, clearance zones, or functional performance goals. Generative approaches have begun to explore text-to-3D city synthesis~\cite{xie2024citydreamer,feng2023layoutgpt} and reconstruction~\cite{10.1145/3757377.3763866,Zhang20TSAS}. However, coupling explicit landscape design rules with quantifiable 3D performance metrics, such as acoustic damping or thermal comfort, remains largely unaddressed.

We present a novel urban landscaping framework for synthesizing heterogeneous vegetation assets. For a given urban layout, our method employs a design-centric approach to generating and optimizing vegetation layout and function. The input is a 2D map of the urban layout and a text prompt describing the layout strategy. Our framework outputs aesthetically plausible vegetation configuration that fulfills high-order functional constraints. We show that modeling urban landscapes can be accomplished by coupling Vision-Language Models (VLMs) for high-level semantic structure with evolutionary optimization (CMA-ES) ~\cite{HansenOstermeier2001CMAES} for geometric refinement. This enables the generation of a broad range of design typologies, such as borders, all\'ees, and garden rooms, while adhering to parcel geometry. Unlike previous models that focus on vegetation location, our framework explicitly evaluates functional performance by estimating shade coverage, sky-view factors (SVF), and acoustic attenuation. \cut{Fig.~\ref{fig:teaser} shows three park designs generated from identical geometric site constraints. Left: a formal layout with regular spacing. Center: a picturesque, asymmetric composition with clustered planting. Right: a modernist scheme characterized by clear, rigorous geometry.}

Our contributions are (1)~a hybrid landscaping framework combining VLM-based structural priors with geometric evolutionary optimization, \ch{(2)~phenomenological functional objectives -- efficient scene-design surrogates for shade, biomass, and acoustic exposure --} and (3) a coupled evaluation scheme for view-dependent aesthetics. \ch{Source code, prompt templates, and configurations are publicly available at \url{https://github.com/greenscapes-project/greenscapes}.}

%% file: src/02-related_work.tex
\section{Related Work}
\label{sec:rw}
Computer graphics research on plant synthesis spans procedural modeling, reconstruction, and interactive authoring.

Early approaches model branching using rule-based systems~\cite{Honda71}, grammars~\cite{Aono:CGaA:84}, particle systems~\cite{10.1145/325334.325250}, and fractals~\cite{Oppenheimer:86}. L-systems~\cite{Pwp:GI86,Prusink1990, Palubicki2019Morphogenesis} provide a principled formalism for plant generation. However, specifying rules for complex trees is labor-intensive and may require learned rule inference~\cite{Guo20ToG,Lee2024Latent}. \cut{To improve usability, procedural modeling methods combine botanical priors with geometric modeling and interactive controls~\cite{10.1145/218380.218427,10.1145/253607.253880,10.5555/2331067.2331083}.} Subsequent work addresses self-organizing growth~\cite{Palubicky:09:SIGG,Runions:05:Sigg,Guo20ToG}, inverse procedural modeling~\cite{Stava2014}, urban forests~\cite{niese2022procedural,10.1145/1944745.1944773}, and environmental response~\cite{Pirk:2012:PTI:2185520.2185546}.

Tree reconstruction has been studied from single or multi-view imagery~\cite{Neubert:07:SIGG,Tan:2008:SIT:1409060.1409061,10.1145/1141911.1141929}, point clouds~\cite{Xu07,Livny:2011:Sigg}, and videos~\cite{Li:2011:MGM:2070781.2024161}. \cut{Several pipelines leverage intermediate representations, such as segmentation masks~\cite{Argudo:2016:SRR:2949067.2949098} or 3D bounding volumes~\cite{Li:2021:SingleImageReconstruction}, to regularize inference of branching and foliage.} \cut{\citet{Bradley:2013:IRS:2461912.2461952} reconstruct dense foliage by combining stereo-derived point clouds with a statistical model.}

Interactive authoring aims to make plant design practical for artists and designers. Sketch-based methods use gestures and silhouettes that guide procedural branching~\cite{OkabeSketchTree06,Tan:2008:SIT:1409060.1409061,10.1145/1409060.1409062}, as well as sketch-driven modeling of plants~\cite{10.1145/1124728.1124746} and flowers~\cite{Ijiri:EG:06}. TreeSketch provides a particularly complete integration of user sketches with procedural generation~\cite{10.5555/2331067.2331083}. 

\begin{figure*}[t]
\centering
\includegraphics[width=1\textwidth]{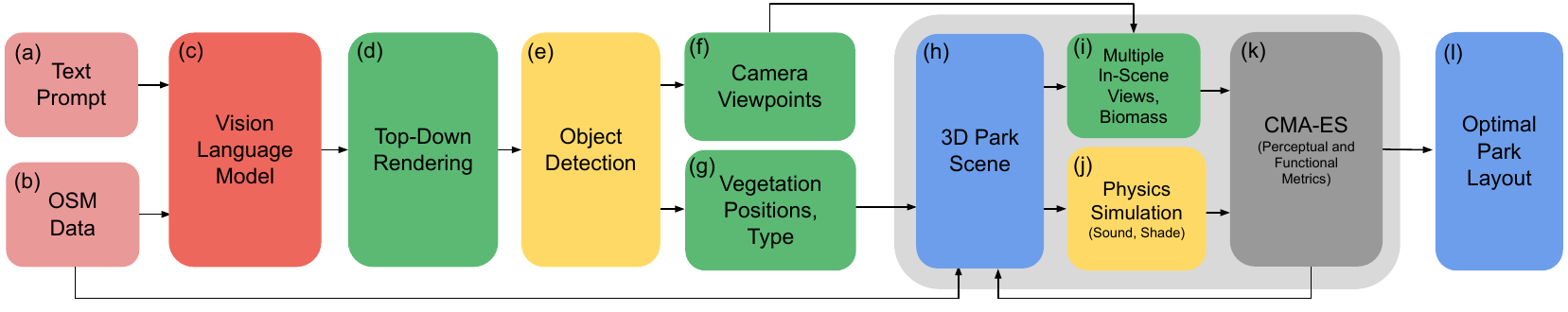}
\vspace{-7mm}
\caption{Overview of our pipeline for prompt-driven park vegetation design. Inputs are a user text prompt (a) and OSM-derived site context (b). An image-to-image constrained editor generates multiple top-down layout proposals (c), which are converted to a metric top-down representation (d) and processed by instance extraction (e) to obtain candidate pedestrian viewpoints (f) and vegetation instance positions and coarse types (g). These instances are projected into world coordinates and assembled into a 3D park scene (h). We evaluate candidate scenes by rendering multiple in-scene views to compute perceptual descriptors and biomass-related quantities (i) and by running physics-inspired field models for functional measures such as shade and sound (j). CMA-ES then refines continuous instance parameters and triggers discrete species substitutions under a Gaussian-mixture asset prior using the combined perceptual and functional objectives (k), yielding the final optimized park layout (l).}
\vspace{-2mm}
\label{fig:overview}
\end{figure*}

Recent generative pipelines have begun to address text- and structure-conditioned synthesis of large outdoor environments by combining layout planning from language and diffusion-based generation of scene representations. On the layout side, LLM and VLM-driven planners generate structured spatial specifications (e.g., bounding boxes, scene elements, or hierarchical layouts) that can be executed by downstream generators, enabling improved controllability and spatial reasoning compared to direct text-to-image generation~\cite{feng2023layoutgpt,Wang_2025_CVPR,Srivastava_2025_ICCV}. For outdoor and urban environments, diffusion models have been extended from object- and indoor-centric settings toward city-scale or real-world outdoor scene synthesis, including semantic scene generation and editing in large outdoor maps~\cite{Lee_2024_CVPR}, text-driven multi-view consistent 3D scene generation that supports long camera trajectories and outdoor content~\cite{Zhang_2024_CVPR}, and coarse-to-fine 3D diffusion schemes designed for detailed large-scale outdoor scenes~\cite{Liu_2024_ECCV}. Structure-conditioned urban generation has explored graph-based formulations, including hierarchical layout generation at the city scale from scene-graph-conditioned diffusion for controllable 3D outdoor scenes~\cite{Liu_2025_ICCV}. LLM-assisted procedural modeling has also been explored for instruction-driven 3D asset construction in 3D-GPT~\cite{Sun2025ThreeDGPT}, text-controlled procedural landscape generation~\cite{Liu2024ControllableLandscapes}, and large-scale procedural scene generation in SceneX~\cite{Zhou2025SceneX}. While these methods substantially improve controllability, they primarily optimize for visual or structural fidelity. In contrast, our approach explicitly couples generative proposals with functional and perceptual evaluation inside an optimization loop.

%% file: src/04-method.tex
\section{Method}\label{sec:method}
The inputs to our pipeline (Fig.~\ref{fig:overview}) are a user-defined text prompt (Fig.~\ref{fig:overview}a) and a lot polygon geometry from OpenStreetMap (OSM)  \cite{HaklayWeber2008OSM} data expressed in a 2D coordinate frame (Fig.~\ref{fig:overview}b). We rasterize the OSM polygon into a top-down conditioning image
and run a \emph{constrained image-to-image generator} (Gemini~3, \cite{gemini3}) instructed to modify only a designated editable region (grass). This produces top-down vegetation-layout proposals (Fig.~\ref{fig:overview}c) that are used as the sole input for class-wise instance extraction (Fig.~\ref{fig:overview}d--e). All experiments use fixed prompt templates rather than interactively tuned prompts. The templates specify the editable mask, the allowed vegetation vocabulary, the required top-down orthographic style, and the requested landscape pattern or design strategy. In our experiments, we used 18 fixed templates: 17 style and layout templates for prompt-conditioned generation and one annotation template for instance extraction. \ch{In parallel, we construct a bounded set of evaluation viewpoints by prompting the VLM to annotate pedestrian standing points and view directions on the conditioning raster (Fig.~\ref{fig:overview}f); the decoding procedure is detailed in the supplementary material.} Detected instances are then \emph{projected} into world coordinates, assigned assets from a library, and assembled into a 3D scene with per-instance attribute vectors and collision proxies (Fig.~\ref{fig:overview}g--h). We evaluate candidate scenes by rendering in-scene buffers from the selected viewpoints and compute image- and simulation-derived metrics (Fig.~\ref{fig:overview}i--j). Finally, we refine continuous per-instance parameters and discrete asset identities using the Covariance Matrix Adaptation Evolution Strategy (CMA-ES) under a low-dimensional parameterization coupled with a Gaussian Mixture Model (GMM) (Fig.~\ref{fig:overview}k). We return the best feasible configuration (Fig.~\ref{fig:overview}l).

\subsection{OSM Preprocessing, Masks, and Assets}\label{sec:method_inputs}
For the target lot and neighborhood context, we query OSM for polygonal geometry (lot boundaries, building footprints, and obstacles) and graph geometry (streets and paths). 

The OSM tags are mapped into a compact set of region classes used throughout the pipeline (e.g., \emph{plots}, \emph{parks}, \emph{alleys}, \emph{urban forest}, and other vegetation-relevant categories). The layers are rasterized into aligned masks and grouped into exclusion and mandatory regions (e.g., building, road, water, pathway, open space, and grass).

We use a library of parameterized vegetation assets \ch{categorized as large trees, small trees, and shrubs}. Each vegetation asset stores an \emph{attribute vector}
\begin{equation}
\mathbf{z} = [\,a,\;b, \;c,\;k_{\mathrm{shade}},\;k_{\mathrm{bio}},\;k_{\mathrm{ac}}\,] \in \mathbb{R}^6,
\label{eq:z}
\end{equation}
where the ellipsoidal bounding volume($\mathcal{E}=(a,b,c)$ is defined by the two horizontal crown radii~$(a,b)$ and its height~$c$, $k_{\mathrm{shade}}$ defines the shadow strength used in the shadow-propagation method~\cite{Palubicky:09:SIGG}, \ch{$k_{\mathrm{bio}}$ linearly maps canopy volume to a biomass proxy~\cite{Jenkins2003Biomass}}, and~$k_{\mathrm{ac}}$ represents acoustic damping and scales a diffusion coefficient of a 3D acoustic diffusion model. The bounding box ellipsoid~$(a,b,c)$ also serves as a Gaussian prior. The values are derived from horticultural size ranges. %

We rasterize the OSM geometry into a top-down orthographic image.
This rasterization defines \ch{a single global axis-aligned affine transform} $T:\mathbb{R}^2\!\rightarrow\!\mathbb{R}^2$ that maps world coordinates (meters) to pixel coordinates. We use~$T$ throughout for all pixel-to-meter conversions.

\begin{figure}[t]
  \centering
  \includegraphics[width=\linewidth]{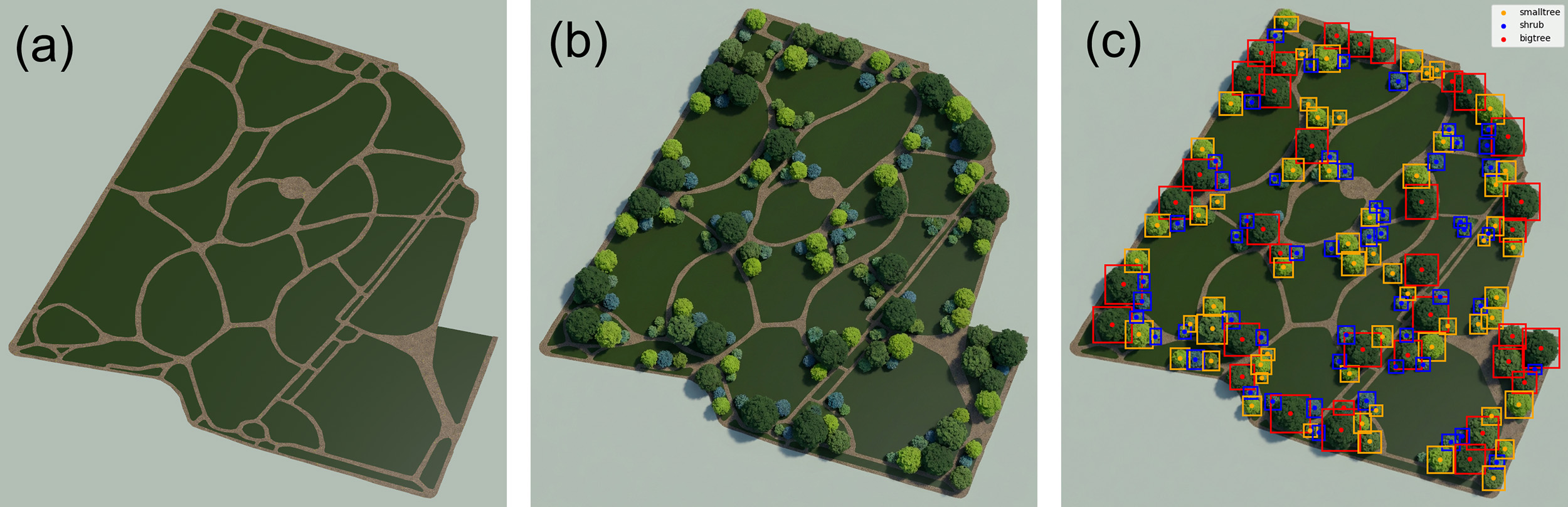}
  \caption{Feasibility and constraint adherence of the first stages of our pipeline. In (a), we show an example conditioning raster obtained from OSM data. In (b), a constrained-edit proposal of the VLM is depicted for that input raster. In (c), we show the VLM-generated bounding box annotation overlay.}
  \vspace{-4mm}
  \label{fig:results_pipeline}
\end{figure}

\subsection{Top-down Layout Generation and Instance Extraction}
\label{sec:method_layout}
We generate vegetation layouts using Gemini 3 Pro~\cite{gemini3} on the conditioning raster, subject to \emph{spatial constraints}: only pixels labeled as grass are editable. Obstacles (paths, buildings, water) and the outside-canvas background are regions where vegetation cannot be placed, and the pixels in these regions must remain unchanged. \ch{The prompt restricts the vocabulary of vegetation assets to (\texttt{largetree}, \texttt{smalltree}, \texttt{shrub}), each represented by a distinct canopy color; lawns form the editable ground region.} Fig.~\ref{fig:results_pipeline}a) illustrates the conditioning raster.

\ch{To accurately recover individual assets from the generated layout, we then perform a second pass with Gemini~3 Pro and prompt the model to return a JSON list of labeled bounding boxes (\texttt{bigtree}, \texttt{smalltree}, \texttt{shrub}) in normalized image coordinates, which we convert to pixel coordinates; for the city-scale examples, a dot-annotation variant is used instead (see supplementary material).} \ch{The extraction} yields an instance list
\(
\pi_i=\{\texttt{class}_i,(u_i,v_i),s_i\}
\)
where $(u_i,v_i)$ is the center of the bounding box and $s_i$ is derived from the box dimensions.

Instance centers are reprojected to world coordinates by the inverse transform $(x_i,y_i)=T^{-1}(u_i,v_i)$. We map each detected class to a vegetation category and select a random asset within that category. Each instance stores the assigned asset identity and its attribute vector $\mathbf{z}$, as defined in Eq.~\eqref{eq:z}. The size proxy~$s_i$ is converted to an initial footprint radius in meters using~$T$ and assigned a canopy ellipsoid~$\mathcal{E}$ with radii~$(a_i,b_i,c_i)$. Continuous per-instance scale parameters are applied to the mesh and ellipsoid.

The VLM output is post-processed deterministically. Each detected bounding box is first assigned to one of the vegetation classes and mapped from image coordinates to lot coordinates by \(T^{-1}\). The box center defines the initial 2D instance position, while the box extent defines an initial footprint that is converted into a 3D bounding ellipsoid. A placement is marked invalid if its center lies outside the editable scene region, if its ellipsoid extends outside the lot, intersects obstacle polygons, overlaps path geometry, or violates vegetation-to-vegetation collision constraints. The invalid placements are repaired in two steps. First, center reprojection moves the instance center to the nearest feasible location on the planting mask. Then, local obstacle correction shifts the bounding ellipsoid along the signed-distance gradient until remaining intersections are removed. The maximum number of repair iterations is 30. If the instance is still invalid, it is discarded. More technical details about the repair step of our method are given in the supplementary material of this paper.

\subsection{In-scene Rendering and Metric Computation}
\label{sec:method_evaluation}
Segmentation buffers for each scene and each selected viewpoint are rendered. \ch{We define eye-level perspective views for skyline shape assessment and a hemispherical fisheye segmentation to compute a sky view factor at all viewpoints generated in the previous stage.} The buffers are inputs to the view-dependent aesthetic metrics. We also compute the scene's biomass based on the biomass of the vegetation assets. Furthermore, we simulate shading and acoustic exposure from the vegetation meshes. \ch{These functional terms are efficient scene-design surrogates rather than calibrated physical predictions.}

\textit{Shade calculation.}
The path-level \emph{direct-sun} shading is estimated using a deterministic \emph{shadow propagation method} from~\cite{Palubicky:09:SIGG}. Monte-Carlo or global-illumination-style estimators are substantially more expensive, while voxel shadow propagation provides a plausible approximation proxy of local light conditions~\cite{NauberMaeder2024LightModels}. 

\textit{Acoustic exposure} is estimated with a two-stage model designed for repeated evaluation during optimization. First, we pre-compute a small set of acoustic descriptors for each vegetation asset from its 3D geometry and acoustic damping value using Monte Carlo ray tracing. Second, we use the descriptors to parameterize a diffusion-with-decay simulation of sound energy on a 3D voxel grid. The resulting exposure is measured at camera view points.

\ch{We trace 20,000 rays for each of 16 incident directions sampled over a hemisphere per vegetation asset in a pre-processing stage.} Each ray performs a single interaction with the geometry. At the intersection point, the outgoing direction is sampled either as a specular or a diffuse reflection, where diffuse directions are sampled from a cosine-weighted lobe. We compute three parameters per asset from the final distribution of outgoing ray directions and energies: (1)~$s_{\mathrm{rand}}$ is the fraction of reflected energy describing random scattering (using ISO-17497-1), (2)~$d_n$ denotes a normalized measure of how spread-out the outgoing energy is over the azimuth, where higher values imply stronger spreading (using ISO-17497-2), and (3)~$\boldsymbol{\mu}$ is the dominant outgoing direction.
\begin{center}
    \includegraphics[width=0.45\columnwidth]{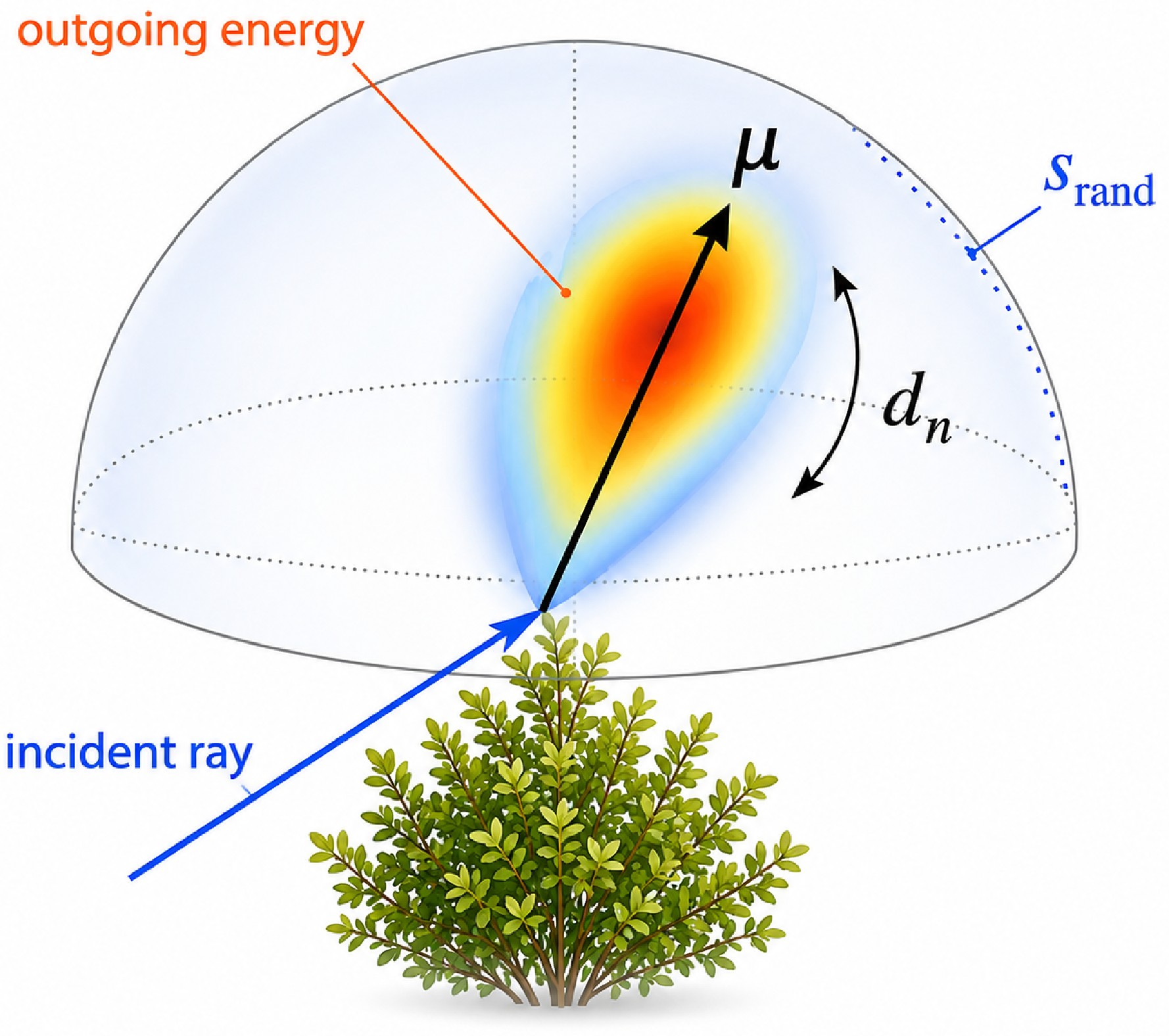}
\end{center}
For reflected rays with outgoing unit directions $\boldsymbol{\omega}_i$ and remaining energies $w_i$, we define
$\boldsymbol{\mu}=\sum_i w_i\boldsymbol{\omega}_i/\|\sum_i w_i\boldsymbol{\omega}_i\|$,
which gives the normalized energy-weighted resultant direction. 

 \textit{Scene-level evaluations.} Given a candidate scene, we discretize the park volume into a 3D voxel grid and label voxels by material class (air, vegetation, buildings, ground). We then simulate the propagation of sound-energy density~$E(\mathbf{x},t)$ with an anisotropic diffusion model with decay and a point-source term. The update is computed with an explicit finite-difference scheme (forward Euler) and a timestep chosen for stability (CFL). The top and side domain boundaries are treated as open space by enforcing~$E{=}0$.

\begin{figure}[t]
  \centering
  \includegraphics[width=\linewidth]{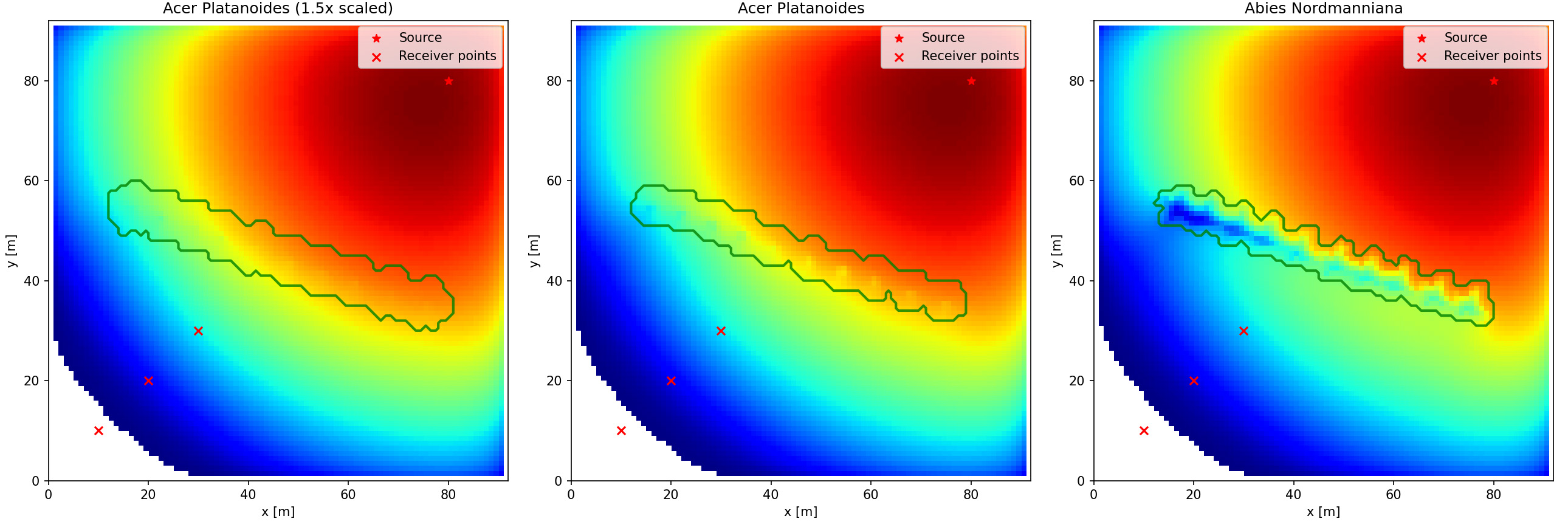}
  \vspace{-4mm}
  \caption{Color-mapped visualization of the ear-height slice of the simulated 3D sound-energy voxel space. A sound source is placed near the upper-right corner (red marker), and a linear tree belt (green outline) separates the source from receiver samples (red markers) below. Increasing the per-vegetation attenuation parameter slows the diffusion update through the tree belt, reducing and delaying energy propagation toward the receivers; the three panels differ only in this attenuation coefficient, while keeping the geometry and discretization fixed.}
  \vspace{-4mm}
  \label{fig:attenuation}
\end{figure}

\begin{figure}[t]
  \centering
  \includegraphics[width=0.49\linewidth]{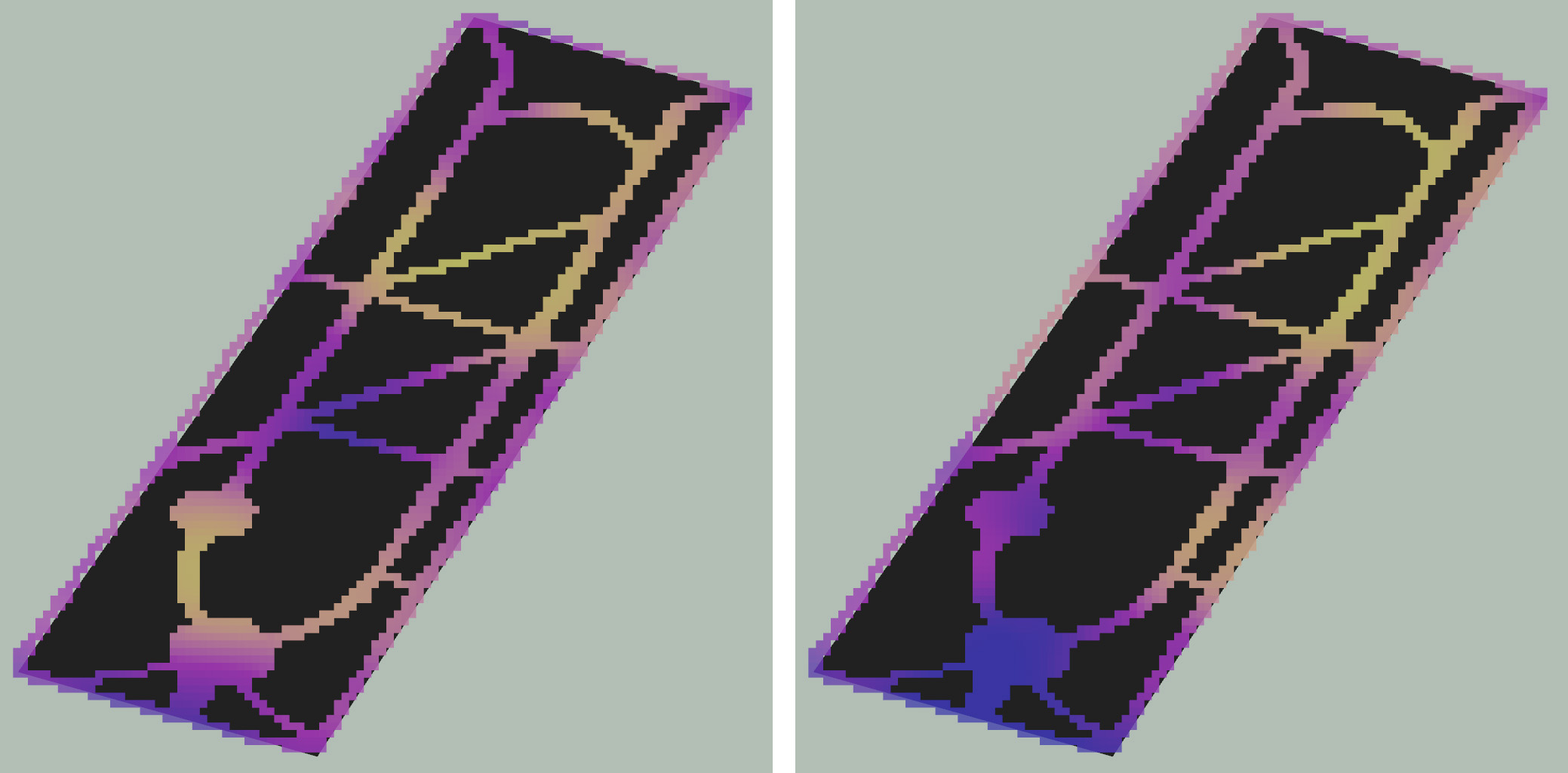}
  \includegraphics[width=0.49\linewidth]{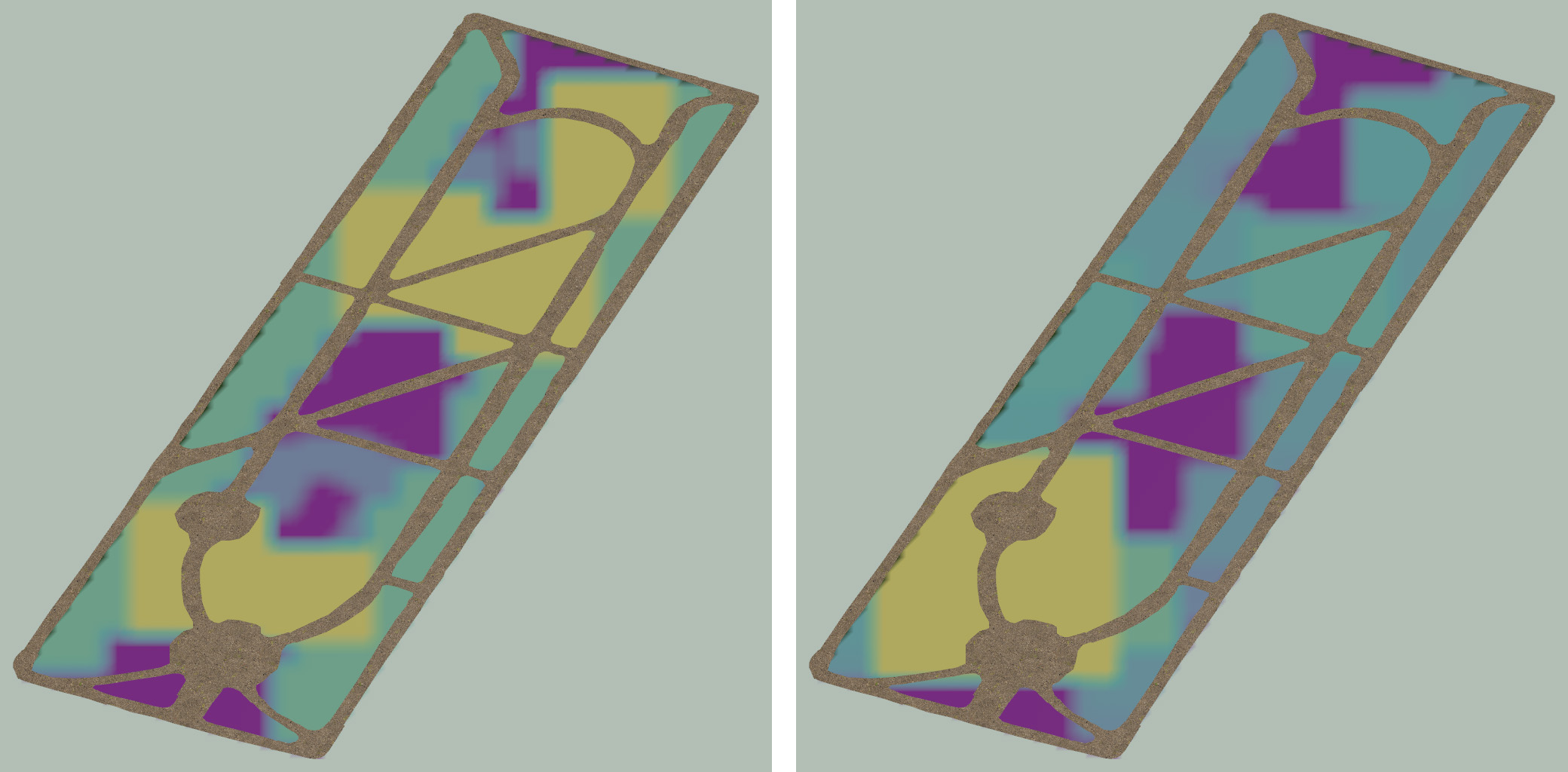}
  \caption{Example park scenes before and after CMA-ES refinement of the VLM prior. Path shading is visualized as a scalar field over the walkable network (plasma colormap), and vegetation biomass is visualized per instance (viridis colormap). Optimization increases functional shading where needed while reallocating plant mass to meet biomass/coverage targets under the same geometric constraints; the images show the resulting redistribution of canopy placement along paths.}  \label{fig:shadow_biomass_opt}\vspace{-6mm}
\end{figure}

We propagate energy density $E(\mathbf{x},t)$ using an anisotropic diffusion equation with decay:
\begin{equation}
\frac{\partial E}{\partial t} \;=\; \nabla\!\cdot\!\left(D(\mathbf{x})\,\nabla E\right)\;-\;\lambda(\mathbf{x})\,E \;+\; S(\mathbf{x},t),
\label{eq:ac_diffusion}
\end{equation}
where $S$ is a source term. Each voxel stores a diffusion tensor~$D$ and a decay rate~$\lambda$ determined by the class of the occupying object (air, vegetation species, building, ground), with vegetation parameters derived from the precomputed asset descriptors.

Vegetation voxels receive anisotropic diffusion and decay parameters from the precomputed descriptors using a fixed global heuristic shared across all experiments. Specifically, higher~$d_n$ increases the base diffusion magnitude as the energy spreads more in space, $\boldsymbol{\mu}$ biases diffusion along a dominant direction (anisotropy), and higher~$s_{\mathrm{rand}}$ increases decay to represent stronger attenuation. Buildings are treated as strong decay regions, while air uses fixed background values. \ch{All numerical mapping constants are listed in the supplementary material.}

We place eight receiver points \ch{at 2\,m ear height, evenly spaced} along the path network and record the simulated energy~$E$ over time. To isolate the effect of vegetation, three simulations are run with identical source, discretization, and boundary handling: one \emph{free-field} reference, one \emph{with} vegetation, and another with higher attenuating vegetation. We report excess attenuation as an energy ratio, expressed in decibels
\begin{equation}
\Delta L_{\mathrm{dB}} = -10 \log_{10}\!\left(\frac{E_{\mathrm{with}}}{E_{\mathrm{free}}}\right),
\end{equation}
visualized as a heat map in Fig.~\ref{fig:attenuation}. This provides a deterministic acoustic objective term comparable across optimization iterations.

\textit{Skyline flatness and periodicity.}
\ch{From the eye-level segmentation view, we compute a binary sky mask.} We then extract a skyline profile~$y(x)$ column-wise as the last sky pixel before the first non-sky pixel (with interpolation over missing columns). Flatness is computed from the detrended skyline (linear trend removed) using three terms: the standard deviation of~$y(x)$, mean absolute first difference, and RMS second difference; these are normalized by a fixed pixel scale and combined into a score $\exp(-(\cdot))$. Periodicity is computed on the same detrended skyline as the average of the dominant-frequency power ratio in the FFT and the strongest autocorrelation peak in a window around the FFT-implied wavelength. \ch{For target-matching experiments, the skyline term measures the deviation of the extracted profile from a target profile, e.g., a sinusoid.}

\subsection{CMA-ES Refinement}\label{sec:method_optimization}
Our main goal is to generate optimal green layouts through non-convex optimization. Therefore, we process the 3D representation of our generated park output through a shade and acoustic exposure simulation pipeline, and compute all relevant skyline and biomass metrics. We then start the final optimization stage of our framework, leveraging CMA-ES. For a selected set of~$M$ CMA-ES scene operators,
\begin{equation}
\theta=\{(s_i^{xy}, s_i^{z}, \mathbf{z}'_i)\}_{i=1}^{M},
\end{equation}
where $(s_i^{xy},s_i^{z})$ are horizontal and vertical scale multipliers and $\mathbf{z}'_i\in\mathbb{R}^6$ is an attribute proposal in the asset space of Eq.~\eqref{eq:z}. \ch{All feasible instances of the repaired configuration are optimized: the VLM proposal fixes the number, class, and 2D position of the instances; the repair operator (Sec.~\ref{sec:method_layout}) runs once before optimization to establish feasibility; and CMA-ES subsequently adjusts only the per-instance scales and attribute proposals, with the GMM assignment triggering discrete species substitutions. Positions are not optimization variables; feasibility during optimization is maintained by clamping the scale parameters to their bounds.}

\begin{figure}[t]
  \centering
  \includegraphics[width=\linewidth]{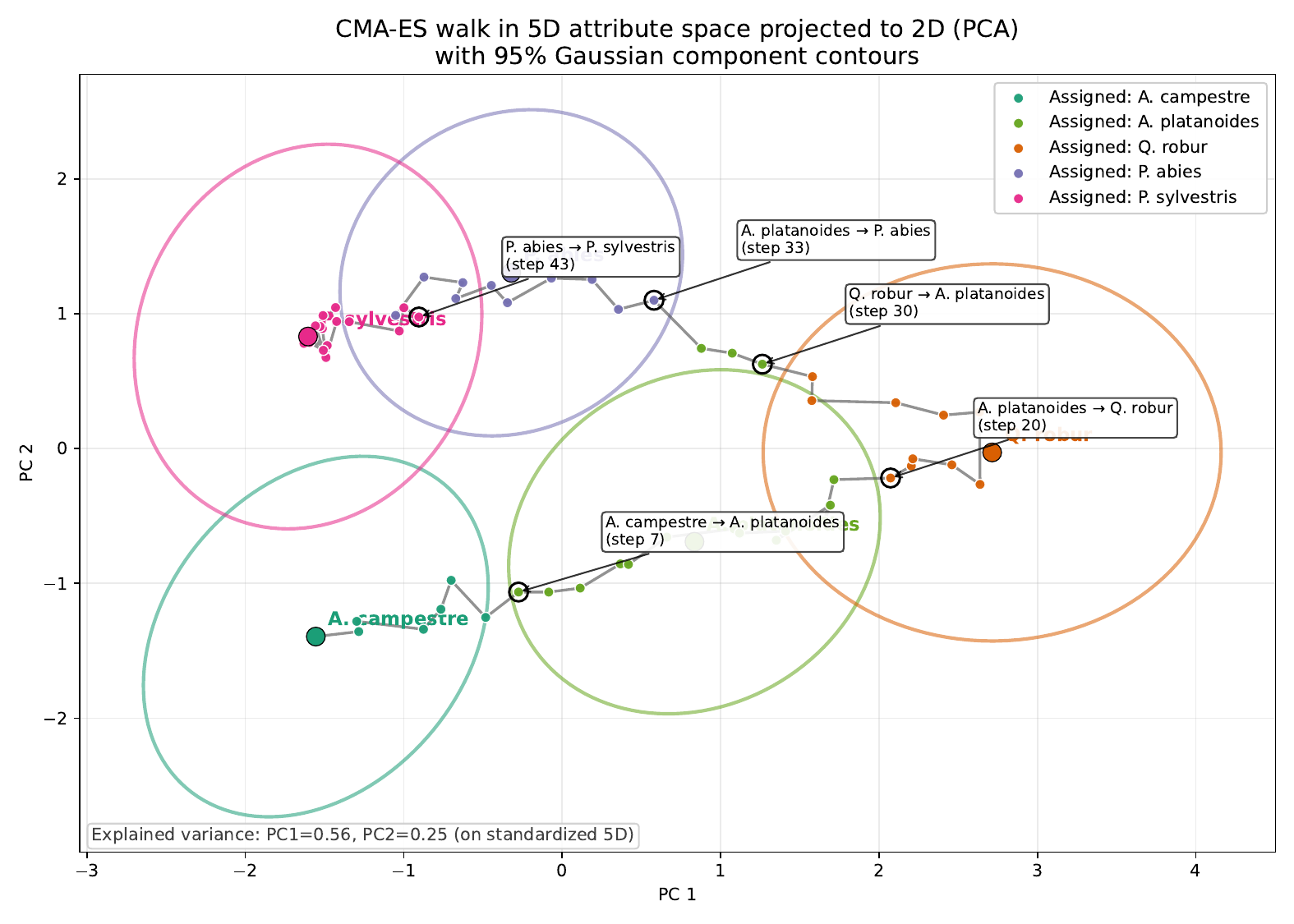}
  \vspace{-7mm}
  \caption{Visualization of the CMA-ES trajectory in the 6D attribute space after PCA projection to 2D. Elliptical contours indicate projected Gaussian components (species priors) used by the GMM-based substitution rule; the polyline shows successive CMA-ES proposals and the induced discrete component assignments.}
  \vspace{-4mm}
  \label{fig:cmaes_walk}
\end{figure}

Each species $j$ defines one Gaussian kernel $\mathcal{N}(\boldsymbol{\mu}_j,\boldsymbol{\Sigma}_j)$ in attribute space (diagonal $\boldsymbol{\Sigma}_j$). Given a proposal $\mathbf{z}'_i$, we assign
\begin{equation}
j^\ast = \arg\max_j \;\pi_j\,\mathcal{N}(\mathbf{z}'_i\mid \boldsymbol{\mu}_j,\boldsymbol{\Sigma}_j),
\end{equation}
and set the discrete asset identity of instance $i$ to $j^\ast$. 
We reject proposals outside an admissible Mahalanobis distance for all assets. In case no valid species is selected, we project~$\mathbf{z}'_i$ back onto the closest ellipsoid defined by the Mahalanobis threshold and use it to set~$(a,b)$ and coefficients~$(k_{\mathrm{shade}},k_{\mathrm{bio}},k_{\mathrm{ac}})$. Fig.~\ref{fig:cmaes_walk} illustrates the resulting \textit{walk-and-jump} behavior. \cut{Initially, an asset starts within the Gaussian kernel that defines the species Acer campestre. As a result of multiple optimization steps, \ch{the asset's attribute proposal} falls within the Mahalanobis threshold for another species (Acer platanoides) and switches its species tag. As a consequence, we replace the asset with another (belonging to Acer platanoides) and continue the CMA-ES optimization loop.}

\begin{figure}[t]
  \centering
  \includegraphics[width=\linewidth]{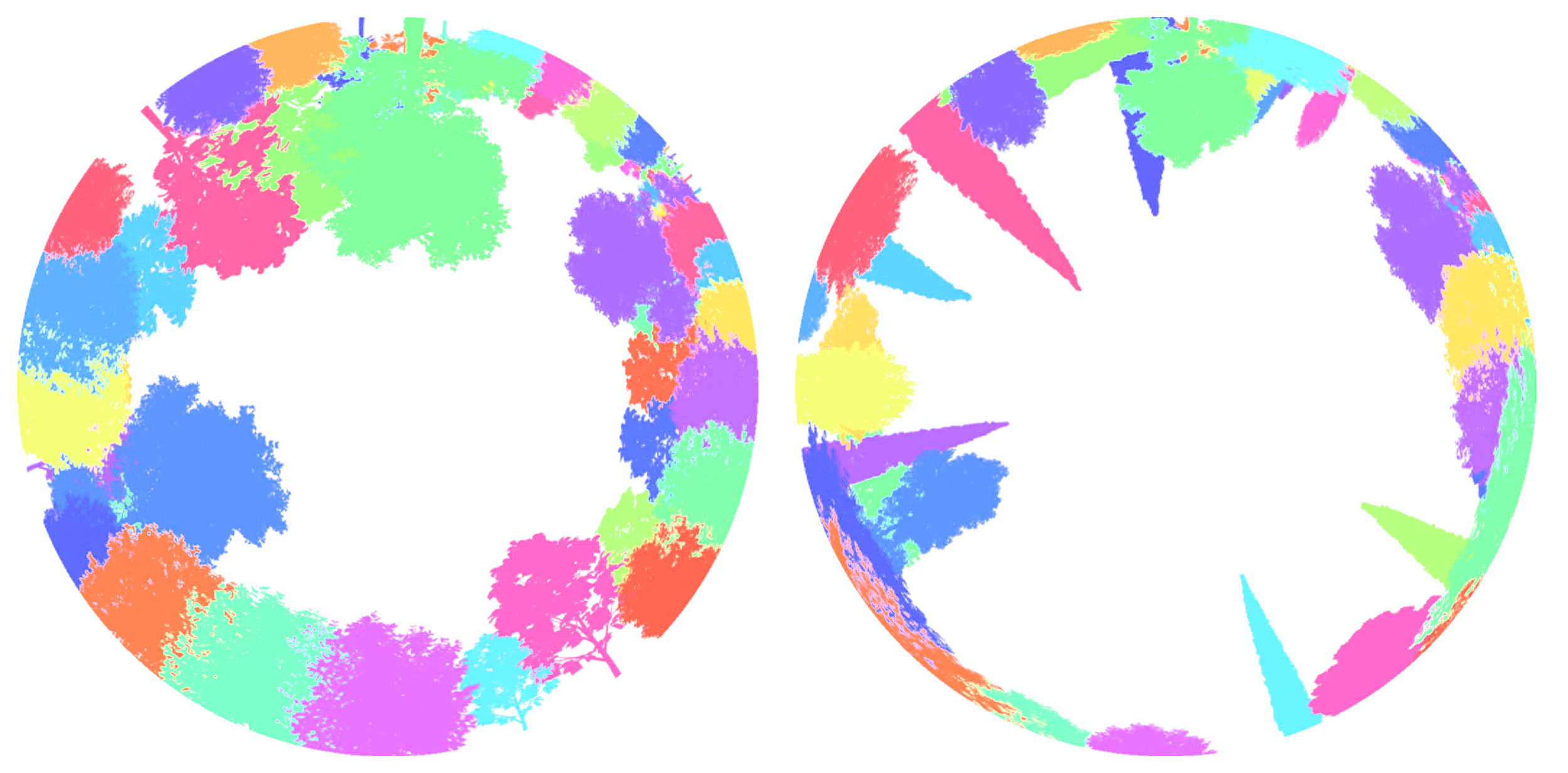}
  \vspace{-6mm}
  \caption{Sky-view exposure before and after CMA-ES refinement from a fixed in-park viewpoint. The fisheye sky segmentation (white = sky) shows increased open-sky visibility after optimization, raising the sky view factor (SVF) to $0.70$ while preserving the surrounding layout constraints.}
  \label{fig:skyview}
  \vspace{-4mm}
\end{figure}

\begin{figure}[t]
  \centering
  \includegraphics[width=\linewidth]{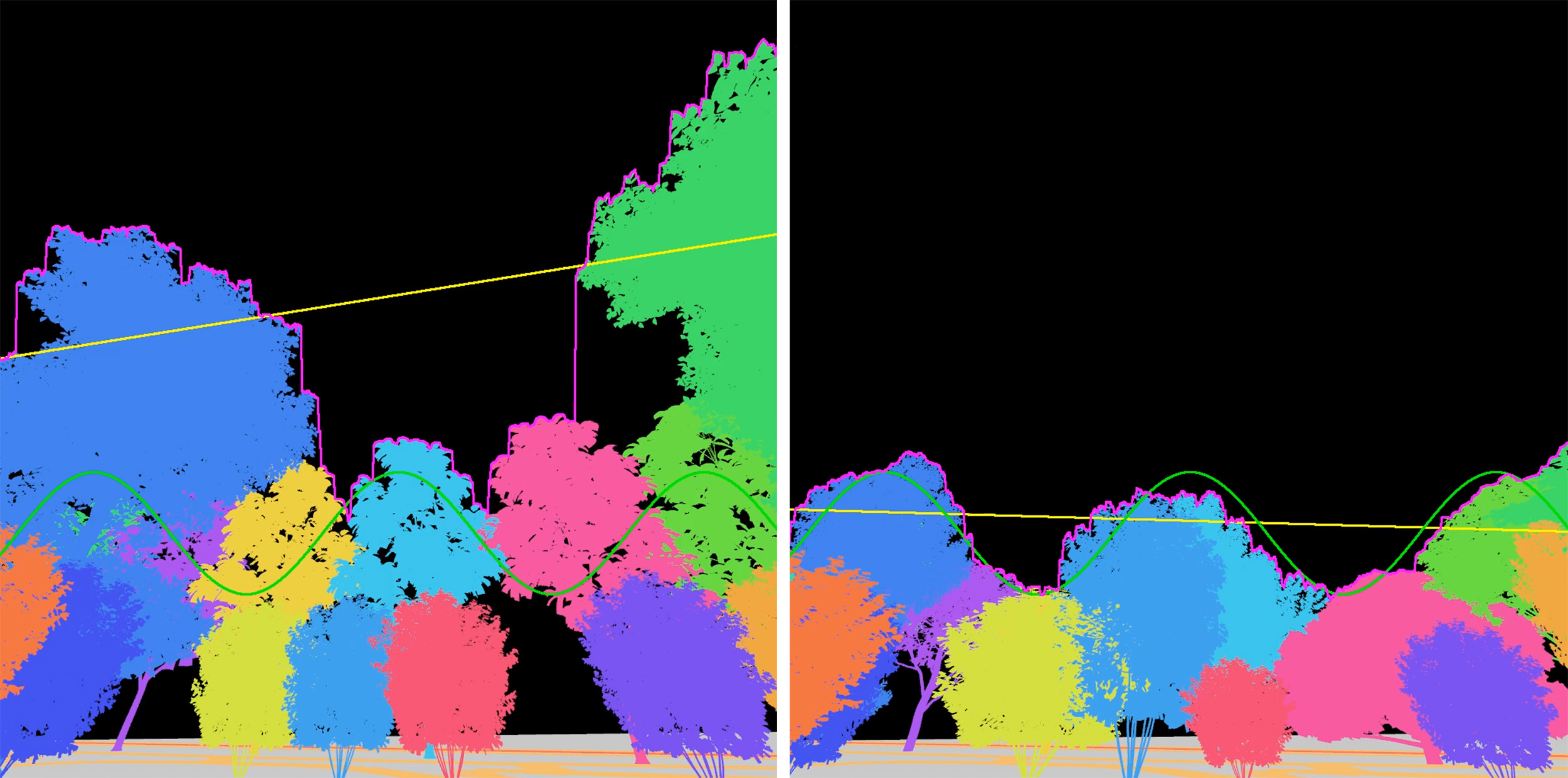}
  \vspace{-6mm}
  \caption{Skyline periodicity matching before (left) and after (right) CMA-ES refinement from a fixed viewpoint. The green curve is the target sinusoidal skyline profile; CMA-ES improves agreement by jointly adjusting per-instance scales and performing discrete species substitutions to introduce canopies with more suitable sizes and crown shapes, yielding a more periodic silhouette while maintaining feasibility constraints.}
  \label{fig:view_sin_opt}
\end{figure}

\ch{Each candidate $\theta$ is evaluated as $E(\theta)$, where $E$ is the weighted sum of five normalized terms -- SVF, a skyline-fit term (flatness, periodicity, or deviation from a target profile, depending on the experiment), path shading, acoustic exposure, and biomass -- plus a small regularizer penalizing deviations from the initial scales. CMA-ES is run with a fixed seed, iteration budget~$G$, initial step size $\sigma_0$, and population size $\lambda_{\mathrm{pop}}=\max\left(4,\left\lfloor 4+3\ln n\right\rfloor\right)$ for search dimensionality $n$; the bounds on~$\left(s_i^{xy},s_i^{z}\right)$ are enforced by clamping. The best feasible configuration over all generations is returned.}

%% file: src/05-results.tex
\section{Results and Validation}\label{sec:results}

\cut{In this section, we present the results of our modeling framework.}

\paragraph{Park-Design Strategies.}
\label{sec:results_embedding}
Figure~\ref{fig:metric_space} shows three design concepts: formal (geometric or axial), picturesque (English landscape), and modernist. \cut{They are commonly treated as distinct and iconic historical styles in landscape design and landscape architecture~\cite{rogers2001landscape}.} To quantify the design principles, we propose orientation entropy and linearity: formal layouts tend to have lower entropy and higher linearity, while picturesque layouts are more irregular.

We embed each scene from its \texttt{GeoJSON} vegetation layout (tree and shrub point locations) by computing two deterministic layout descriptors in a common metric frame (EPSG:3857): \emph{orientation entropy} is the Shannon entropy of a 36-bin histogram of nearest-neighbor bearing angles modulo $180^\circ$ (low for axis-aligned grids; high for irregular and circulinear layouts), while \emph{linearity} is the fraction of points whose two nearest neighbors are approximately collinear (within $5^\circ$), reported as a percentage. We plot 33 reconstructed reference parks with known design strategies and 30 generated parks from three OSM sites and style-specific prompts (three templates per style, with LLM-generated textual variants); the generated parks are rendered as circles. The clusterings show that the two descriptors separate the three strategies and that prompt-conditioned generations align with the corresponding reference distributions, supporting both metric validity and reproducible style control from the \texttt{GeoJSON} layouts.

\paragraph{Prompt-conditioned Controllability.}
\label{sec:results_controllability}
Figure~\ref{fig:styles} shows results with our 3D generative pipeline \ch{from a single OSM input}, but different text prompts. We use the parcel boundary and OSM-derived path network of St.\ Mary's Park and reconstruct a reference vegetation layout from the NYC Tree Map inventory (geolocated trees with species and size attributes)~\cite{NYCTreeMap}. Starting from the single OSM masks, we vary only the text prompt (e.g., \emph{agroecological}, \emph{East Asian}, \emph{infrastructural}, \emph{post-modernist}, \emph{formal}, \emph{picturesque}, \emph{modernist}, \emph{memorial}), yielding diverse 3D layouts that appear as plausible design alternatives to the reconstructed St.\ Mary's Park baseline (top left panel in Figure~\ref{fig:styles}). The results show generated complex vegetation layouts\ch{; e.g., the \emph{agroecological} layout features} haphazardly distributed trees and bushes, whereas the \emph{modernist} and \emph{memorial} layouts are sparser and more organized. 

\paragraph{CMA-ES Refinement.}
\label{sec:results_cmaes}
Figure~\ref{fig:Joyce} shows an objective-ablation study on Joyce Kilmer Park. Panel (a) shows the reconstructed reference layout \ch{from the NYC tree inventory (3D trees placed at their real locations)}, used as the input condition. Panels (b--f) show five optimization configurations in which one objective term is assigned a dominant weight while \ch{the remaining terms are assigned low or zero weights}. \cut{This protocol isolates the qualitative effect of each criterion while keeping the site boundary, path network, feasibility constraints, asset library, and optimizer fixed.} A key observation is that CMA-ES can substantially change the perceived park design without changing the underlying planting positions. Most of the visible differences arise from continuous rescaling and GMM-driven discrete substitutions (crown width, height propensity, and species-level density proxies). The skyline-targeted run (b) exhibits a distinct, non-local signature: alternating canopy heights emerge along the view direction to match a periodic skyline profile. \emph{Increasing SVF} (c) results in species with tall or wide crowns being replaced by slimmer forms so that the same spatial arrangement gives the impression of a more open, plaza-like room from interior viewpoints. \emph{Minimizing shade} (d) selects narrower, more open crowns and lower effective canopy density. \emph{The noise-reduction target} in~(e) yields a belt-like planting appearance by selecting species with higher attenuation for regions near the viewpoints. Finally, \emph{maximizing biomass} (f) increases visual mass by selecting broader-canopy assets and higher scale factors, yielding a denser canopy and reduced long-range sightlines. \ch{The exact weights and targets for each experiment are listed in the supplementary material and released with the code.}

\paragraph{Planting Layouts of Lots.}
Figure~\ref{fig:niese_comparison} provides a qualitative SOTA comparison to Figure 6 in the procedural urban-forestry method of \citet{niese2022procedural}. \cut{Their method defines representative procedural planting strategies, including semi-random, boundary-biased, and clustered placement.} Our prompt-conditioned pipeline qualitatively reproduces these baseline planting patterns.

Figure~\ref{fig:lots} demonstrates this capability on multiple reconstructed site lots derived from OSM. For each example, we use the OSM parcel boundary and building footprint polygons, extrude the footprints into 3D building masses (blue), and treat these as exclusion constraints for vegetation placement. Despite substantial variation in parcel shape, footprint size, and building placement, the generated layouts remain constraint-adherent and adapt the planting structure to the remaining ground area\cut{: vegetation concentrates along parcel edges and setbacks, fills residual wedges and corners, and forms strip-like bands where the geometry induces narrow corridors}.

\paragraph{Runtime Analysis.} The system was implemented in Rust using Bevy for rendering and CUDA for acoustic pre-processing. All results were generated on a laptop running Windows 11 and equipped with an Intel Core i9-14900HX CPU (24 cores), 32\,GB of RAM, and an NVIDIA GeForce RTX 4080 mobile GPU. The average wall-clock time for one optimization generation (processing the full population of candidate scenes) is approximately $23$ seconds. Consequently, a typical full experimental run of $100$ generations requires approximately $38$ minutes to converge. The initialization phase (VLM generation and detection) incurs a fixed one-time latency of $\sim 15$--$20$ seconds depending on API response times.

\paragraph{Quantitative Validation and Robustness}
\label{sec:results_quantitative}
Table~\ref{tab:ablation_timing} summarizes $N=30$ independent runs \ch{with independently sampled VLM layouts (the VLM API is unseeded; downstream stages use fixed seeds)} and reports absolute deviations from the optimization targets. The reconstructed park layout yields a high total cost and fails to meet several prompt-specific objectives, such as SVF and Acoustics. The VLM prior exhibits substantial \ch{sample} sensitivity, with large run-to-run variance in several terms (e.g. skyline-related error), whereas the CMA-ES refinement consistently reduces both the mean target error and the variance across runs, indicating that the optimization stage significantly improves objective alignment over the VLM prior.

To isolate the role of the VLM prior, we additionally perform a component ablation on Joyce Kilmer Park (Table~\ref{tab:component_ablation}). We compare random-position initialization, random initialization followed by CMA-ES, the VLM-derived prior, and the final VLM-initialized CMA-ES result. The VLM prior already improves shade scores over random placement while keeping biomass in a comparable range, indicating that the prompt-conditioned layout provides a useful semantic initialization rather than only random feasible samples. Starting from random placement, CMA-ES improves the functional scores, but remains less effective than VLM-initialized CMA-ES because random layouts often place vegetation away from paths. The final VLM+CMA-ES result achieves the highest shade and acoustic scores while minimizing biomass. 

\begin{table}[t]
    \centering
    \caption{Quantitative error analysis over $N=30$ experimental runs reports the \textit{mean absolute deviation from design targets} ($\mu \pm \sigma$) to demonstrate robustness. While the \emph{VLM Prior} exhibits high variance, the \emph{Optimized} result shows significantly lower standard deviation, indicating that our method reliably converges to the design goals regardless of initialization. Real Park is a fixed reference geometry.}
    \vspace{-4mm}
    \label{tab:ablation_timing}
    \scalebox{0.9}{
    \begin{tabular}{lccc}
        \hline
        Metric (Target) & Real Park (Ref) $|\Delta|$ & VLM Prior $|\Delta|$ & Optimized $|\Delta|$ \\
        \hline
        SVF ($0.75$)         & $0.12$        & $0.06 \pm 0.04$        & $\mathbf{0.03 \pm 0.01}$        \\
        Shade ($2.5$)        & $0.32$        & $0.80 \pm 0.20$        & $\mathbf{0.28 \pm 0.07}$        \\
        Acoustics ($3.5$)    & $5.76$        & $0.50 \pm 0.14$        & $\mathbf{0.21 \pm 0.06}$        \\
        Biomass ($75k$)      & $9{,}736$     & $18{,}683 \pm 6{,}174$ & $\mathbf{896 \pm 190}$          \\
        Skyline Fit ($10k$)  & $10{,}213$    & $9{,}433 \pm 1{,}815$  & $\mathbf{2{,}986 \pm 456}$      \\
        Total Cost           & $1.30$        & $0.40 \pm 0.10$        & $\mathbf{0.05 \pm 0.01}$        \\
        \hline
    \end{tabular}
    }  \vspace{-4mm}
\end{table}

\begin{table}[h]
  \centering

  \caption{Ablation of the two components of \ch{our method: initialization} (random vs.\ VLM-derived prior) and CMA-ES refinement (on vs.\ off)-reported on Shade, Biomass, and the optimization \emph{Cost}, the weighted squared distance to the Shade and Biomass targets (lower is better). CMA-ES on the VLM prior achieves the lowest cost ($0.006$), reaching the shade target ($2.04$ vs.\ $2.00$) while keeping biomass within $\sim$2.3\% of baseline; CMA-ES on random initialization falls short of the shade target ($1.91$) and overshoots biomass by $\sim 5.5\%$, ending $\sim 20\times$ further from the joint target.}
  \scalebox{0.9}{
  \label{tab:component_ablation}
  \begin{tabular}{lccc}
    \toprule
    Run & Shade & Biomass & Cost \\
    \midrule
    Target & 2.00 & 56,317 & 0.0000 \\
    Random Init. & 1.65 & 54,473 & 0.2012 \\
    Random+CMA-ES & $1.91 \pm 0.04$ & $59,447 \pm 1,176$ & $0.1205 \pm 0.0075$ \\
    VLM Prior & 1.70 & 56,317 & 0.0227 \\
    VLM+CMA-ES & $2.04 \pm 0.06$ & $57,581 \pm 780$ & $0.0059 \pm 0.0030$ \\
    \bottomrule
  \end{tabular}
\vspace{-6mm}

   }
\end{table}

%% file: src/06-conclusion.tex
\section{Discussion and Limitations}
Generating plausible vegetation placements in urban scenes is challenging because solutions must simultaneously satisfy geometric constraints and align with human expectations for planting structure and landscape style. Prior work tends to restrict the design space; for example, \citet{niese2022procedural} \ch{focuses primarily on trees and does not model understory vegetation, nor does it explicitly couple functional performance with perceptual architectural strategies; in contrast, our method configures heterogeneous multi-class plantings under such coupled objectives.}

A core bottleneck is that many landscape-architectural principles are qualitative and therefore difficult to formalize in a procedural manner. Our pipeline separates responsibilities accordingly: the constrained image-edit prior provides stylistically coherent, human-plausible structure, while CMA-ES operates on a low-dimensional 3D parameterization to optimize scene-derived objectives via statistical evaluation of in-scene renderings and \ch{physics-inspired surrogate metrics}. However, this gain in expressive power of scene modeling comes at the cost of algorithm runtime. 
Another limitation is that our functional metrics are designed as optimization surrogates, approximative in nature, rather than first-principles-based physical simulations. Finally, our method has a slower runtime than procedural urban vegetation configuration approaches, which report placing 50K trees per minute (\cite{niese2022procedural}) but do not optimize for functional or aesthetic objectives.

\section{Conclusion}
We introduced a pipeline for urban vegetation design that combines a constrained image-edit generative prior with a low-dimensional, simulation-driven CMA-ES refinement loop. The generative stage produces stylistically coherent top-down layouts that capture perceptual landscape strategies, while the optimization stage improves functional and view-dependent objectives by performing continuous \ch{scale adjustments} and discrete asset substitutions within an attribute-space GMM. Our results show prompt-conditioned controllability, metric-consistent clustering with reconstructed exemplars of known styles, and substantial refinement gains beyond the generative prior, suggesting a practical route to integrate high-level aesthetic intent with \ch{surrogate-based} performance criteria in automated park design.

%% file: src/07-figures.tex
\begin{figure*}[htbp]
  \centering
  \includegraphics[width=1.0\linewidth]{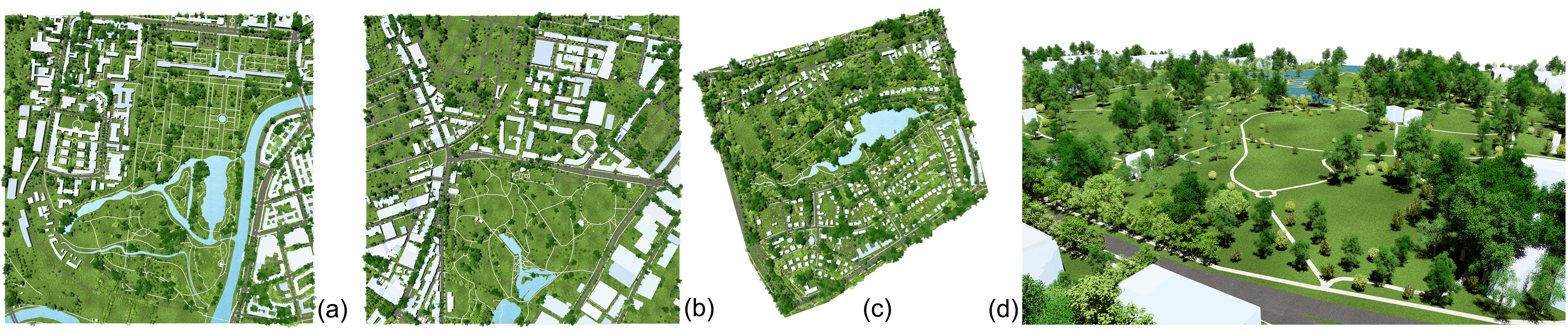}
  \vspace{-8mm}
  \caption{
City-scale vegetation placement examples generated from OSM-derived urban context.
Panels show the area around (a) Sydney Park, Australia,
(b) the Mausoleum Charlottenburg area in Berlin, Germany,
and (c) the Sołacz district in Poznań, Poland, with over 4000 placed tree instances per scene.
Panel (d) shows an in-scene rendering of a generated layout.}
  \label{fig:parks}
\vspace{1mm}
  \centering
  \includegraphics[width=1.0\linewidth]{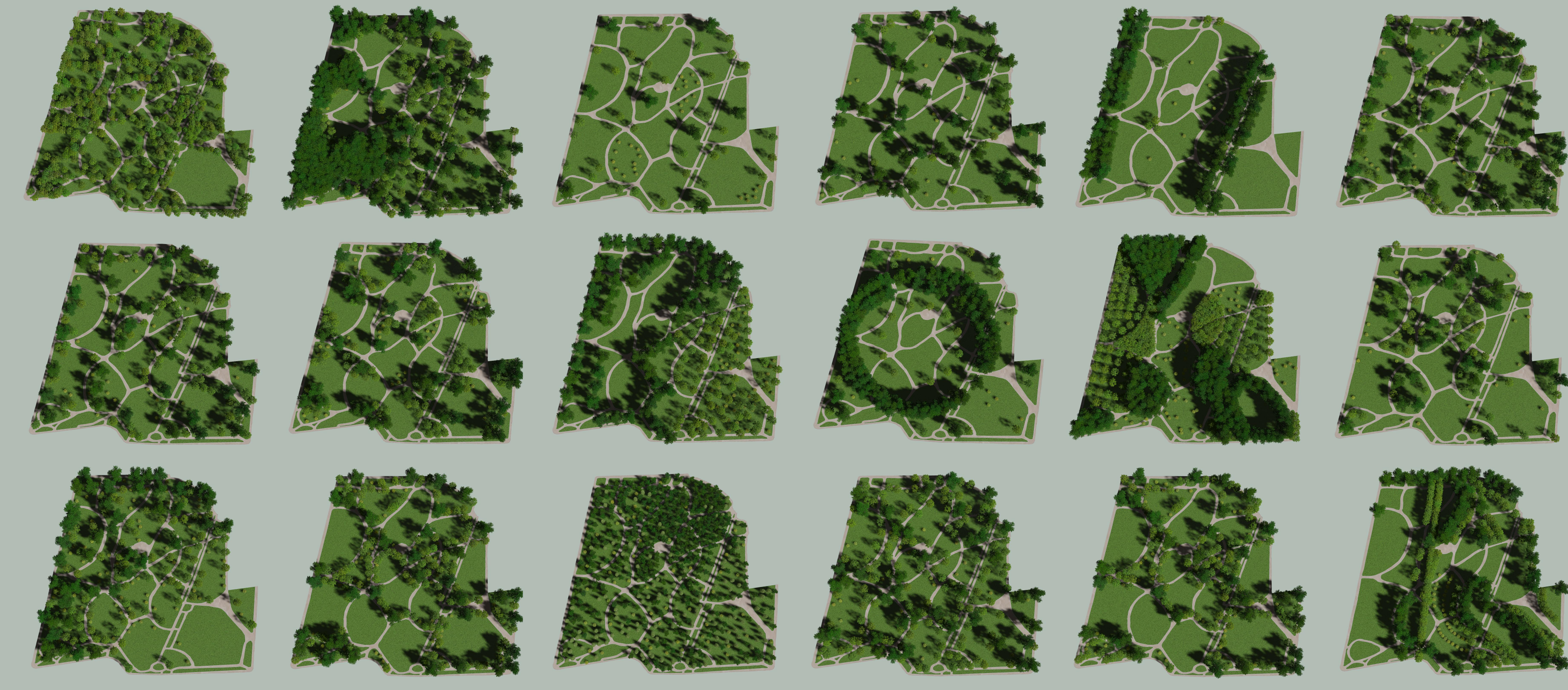}
  \vspace{-7mm}
  \caption{The top left panel is a reconstruction of vegetation positions of St. Mary's park in New York. All other panels show the final 3D layout produced by the reconstructed 3D scene from VLM prior. Varying only the text prompt yields diverse, style-consistent compositions including \emph{agroecological}, \emph{East Asian}, \emph{infrastructural}, \emph{post-modernist}, \emph{formal}, \emph{picturesque}, \emph{modernist}, and \emph{memorial}.}
  \label{fig:styles}
\vspace{2mm}
  \centering
  \vspace{1mm}
  \includegraphics[width=\linewidth]{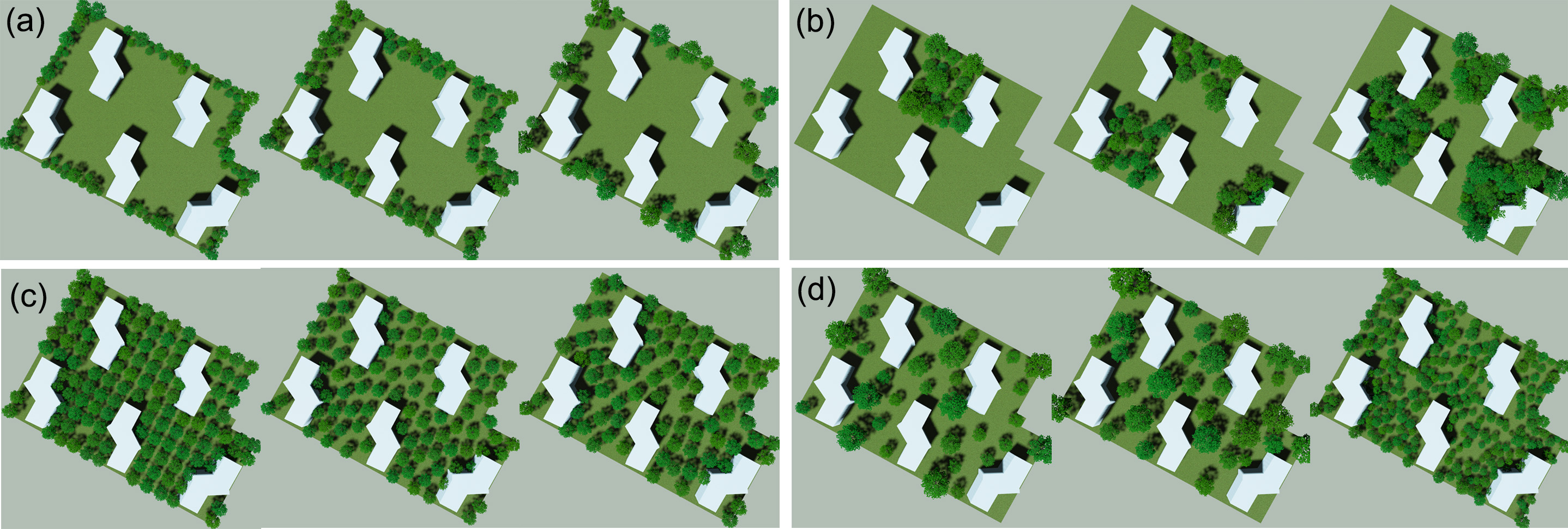}
  \vspace{-7mm}
  \caption{All panels show outputs generated by our method using text prompts that encode the vegetation-placement patterns illustrated in Fig.~6 ~\cite{niese2022procedural} showing: semi-random placement, where trees are sampled throughout the active lot region; boundary placement, where samples are concentrated in a band near the lot boundary; and cluster placement, where samples are concentrated around local cluster regions. In the reference method, these strategies are implemented with Variable Radii Poisson-Disk Sampling.}
  \label{fig:niese_comparison}
\end{figure*}

\begin{figure*}[t]
  \centering
  \vspace{3mm}
  \includegraphics[width=1.0\linewidth]{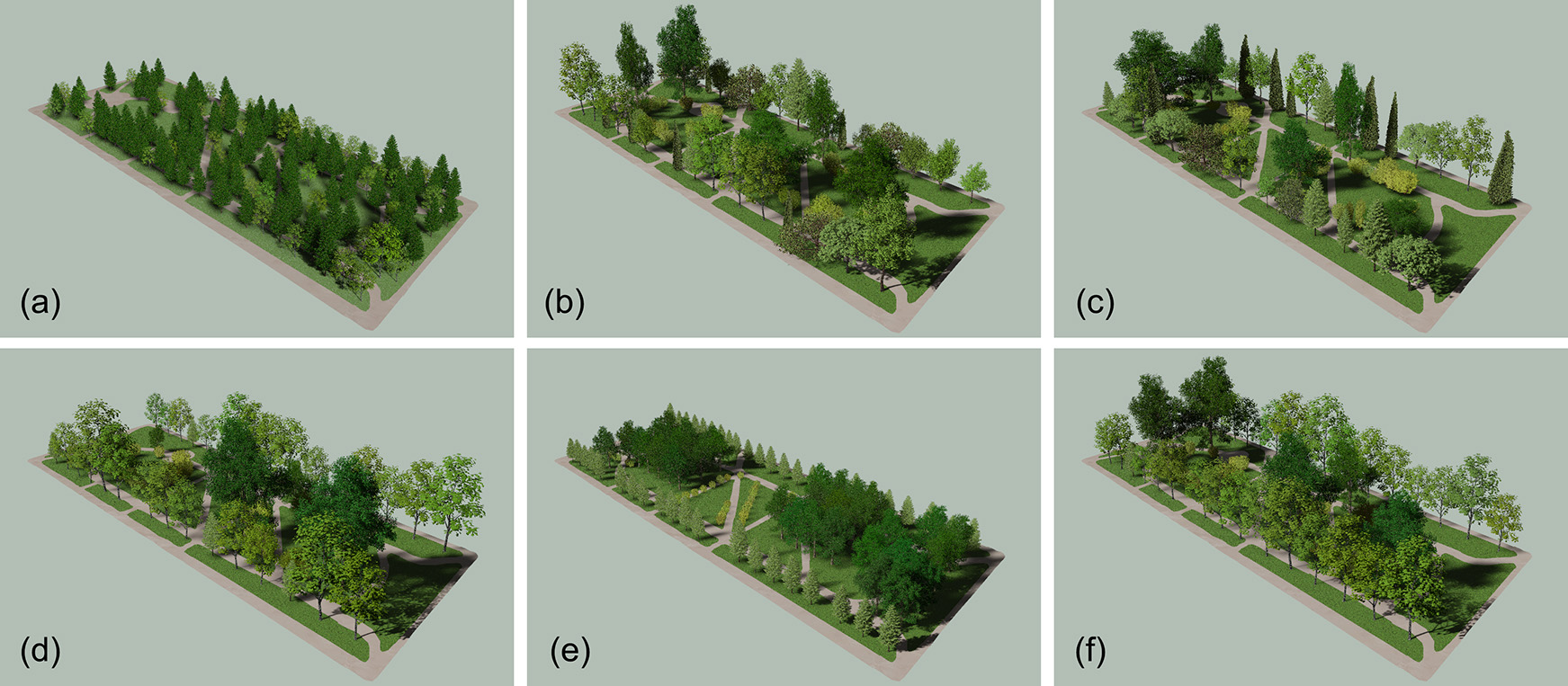}
  \vspace{-3mm}
  \caption{Objective-ablation results for Joyce Kilmer Park. Panel~(a) shows the reconstructed reference layout used as the input condition. Panels~(b--f) show CMA-ES refinements obtained by assigning one objective term a dominant weight while keeping \ch{the remaining terms at low or zero weights}: (b) skyline sine-function fit, (c) SVF ratio, (d) minimal path shading, (e) acoustic/noise-reduction objective, and (f) maximum biomass. All runs use the same site boundary, path network, feasibility constraints, asset library, and optimizer settings; the visual differences therefore illustrate the qualitative effect of each dominant objective.}
  \label{fig:Joyce}
\end{figure*}

\begin{figure}[htbp]
  \centering
  \vspace{4mm}
  \includegraphics[width=\linewidth]{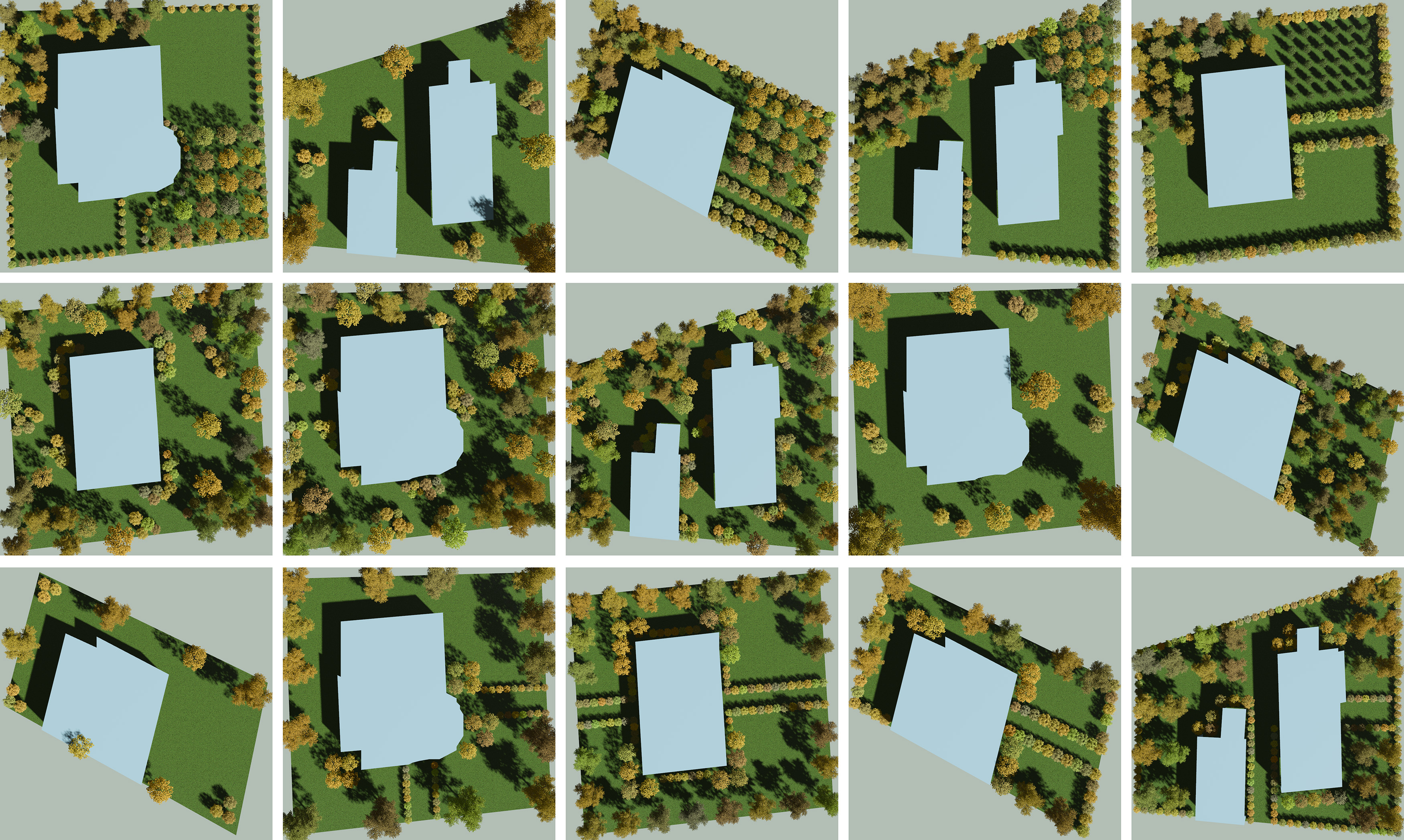}
  \caption{Examples of generated planting layouts for different site lots. For each lot geometry, our method produces distinct plant configurations, demonstrating its ability to adapt planting structure and spatial organization to varying parcel shapes and constraints.}
  \vspace{30mm}
  \label{fig:lots}
\end{figure}

\begin{figure}[htbp]
  \centering
  \vspace{4mm}
  \includegraphics[width=\linewidth]{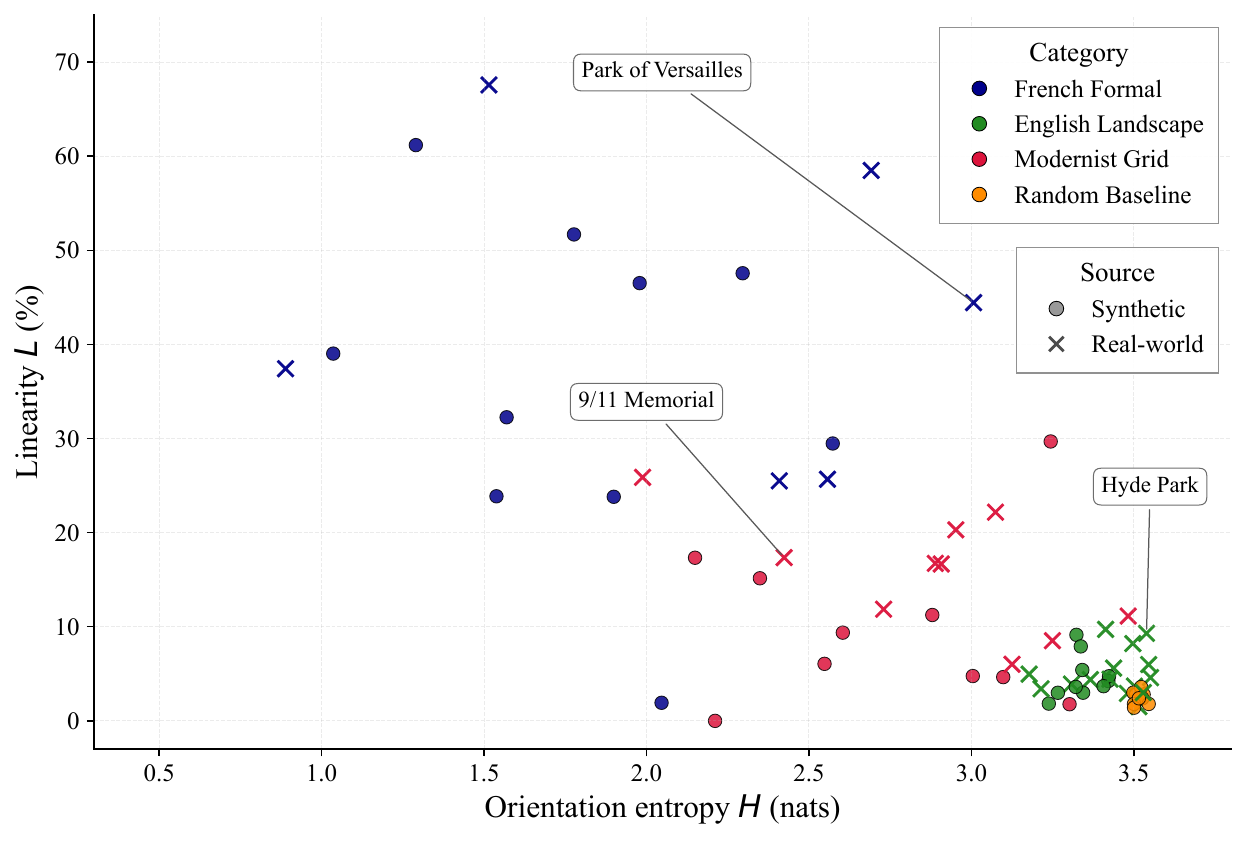}
  \caption{We plot $33$ reconstructed reference parks with known architectural styles and $30$ generated parks produced from OSM sites and style prompts. Colors indicate the intended style (\emph{formal} blue, \emph{picturesque} green, \emph{modernist} red). Disks versus crosses distinguish generated from reconstructed references. The resulting clustering shows that the proposed entropy and linearity metrics separate the three strategies and that the generated parks align with the corresponding reference distributions.}
  \label{fig:metric_space}
  \vspace{12mm}
\end{figure}